\documentclass[aps,prd,reprint,onecolumn,notitlepage,superscriptaddress]{revtex4-1}

\usepackage{lipsum}
\usepackage{amsmath}
\usepackage{graphicx}
\usepackage{color}
\usepackage{tikz}
\usepackage{verbatim}

\usepackage{amssymb}
\usepackage{bm}

\newcommand{\rom}[1]{\textup{\uppercase\expandafter{\romannumeral#1}}}

\begin{document}

\title{Hamiltonian Formalism for  Nonlocal Gravity Models}
\author{Pawan Joshi\footnote{email:pawanjoshi697@iiserb.ac.in}
       }

\affiliation{Indian Institute of Science Education and Research Bhopal,\\ Bhopal 462066, India}

\author{Utkarsh Kumar\footnote{email:kumaru@ariel.ac.il}
        }
        \affiliation{Department of Physics, Ariel University, Ariel 40700, Israel }
\author{Sukanta Panda \footnote{email:sukanta@iiserb.ac.in}
       }
\affiliation{Indian Institute of Science Education and Research Bhopal,\\ Bhopal 462066, India}



\begin{abstract}
Nonlocal gravity models are constructed to explain  the current acceleration of the universe. These models are inspired by the infrared correction appearing in Einstein Hilbert action. Here we develop the Hamiltonian formalism of a nonlocal model by considering only terms to quadratic order in Ricci tensor and Ricci scalar. We also show how to count degree of freedom using Hamiltonian formalism in this model.

\end{abstract}

\maketitle
\section{Introduction}
It is certain that universe has entered into an accelerated phase of expansion recently. 
The simplest theoretical explanation of accelerated expansion is provided by introducing a cosmological constant in the Einstein-Hilbert action. 
However, the addition of such a constant creates a fine tuning problem in the theory\cite{RevModPhys.61.1}. Another way to solve this problem by 
the presence of nonlocal terms(without introducing cosmological constant) 
which appear in the Einstein action by taking ultraviolet(UV) and infrared(IR) corrections into account.
From cosmological model building point of view non-local gravity models 
have attained various achievements such as (1) it can be employed to study 
the cosmology in both IR and UV region, (2) it has a well behaved cosmological 
perturbation theory and (3) the resulting cosmology fits well with 
most of the cosmological data. 

An effective IR corrected Nonlocal gravity model was first developed by Wetterich\cite{Wetterich:1997bz} where correction terms like $ R\Box^{-1}R $ appeared in Einstein Hilbert action. It is found that it doesn't give rise to correct background cosmological evolution. After a decade Deser and Woodard \cite{Deser:2007jk} (Nonlocal Cosmology) introduces a non local model of the form $ Rf(\frac{1}{\Box}R) $ along with Einstein Hibert action. One can describe a correct background cosmological evolution with a certain form of $f(\frac{R}{\Box}).$ Other works in this direction can be found in ref.\cite{Barvinsky:2002uf, Barvinsky:2011hd, Park:2017zls,Kumar:2018pkb,Kumar:2019uwi,Kumar:2019lzp}. 

It is well known that higher derivative theories of gravity contain ghost in their spectrum. Ghost appears in their classical and quantum version. This can be deduced from studying graviton propagator in these theories.  There are two ways to study the occurrence of  ghost in these theories. First way to study the Lagrangian of the system under study and second way to write the Hamiltonian of the system. From the Hamiltonian point of view we look for two points, i.e, first it's boundedness from below and second is the appearance of Ostrogradsky instability\cite{Label1} due to the presence finite number of derivatives higher than two. It is also shown that higher derivative theory of gravity can be renormalizable but only at the cost of unitarity\cite{Stelle:1976gc, VanNieuwenhuizen:1973fi}. Whereas Ostrogradski instability is not an issue for certain UV corrected non-local models containing infinite number of derivatives as it avoids the issue of ghosts and  recovers general relativity (GR) at low energies \cite{Biswas:2011ar, Mazumdar:2017kxr, Biswas:2013kla}. As GR is a theory of space time diffeomorphism invariant that the action can contain all possible diffeomorphism invariant term admitting higher derivatives or infinite derivatives of Ricci scalar, Ricci tensor and Weyl tensor. 

  For a general discussion of the Hamiltonian formalism for the constraint system, refer to \cite{Wipf:1993xg, Dirac:1958sq, Dirac:1958sc, Dirac1, Matschull:1996up}. For UV corrected nonlocal models, such a formalism has been carried out in detail for a nonlocal Lagrangian containing infinite higher derivative terms of Ricci scalar, Ricci tensor and Weyl tensor \cite{Biswas:2013cha, Biswas:2016etb, Biswas:2016egy, Biswas:2012bp, Talaganis:2014ida}. In this paper we carry out a Hamiltonian formulation for a toy model with inverse Laplacian operators acting on Ricci scalar and Ricci tensors. We perform a general Hamiltonian formalism in the equivalent scalar tensor form of our theory. We use ADM formalism to separate out spatial and time derivative terms in Lagrangian.  Primary and secondary constraints are found and from their Poisson brackets, we obtain the first and second class constraint. This, in turn, helps us to count the number of degree of freedom. Such a formulation for a model only with Ricci scalars has been studied in \cite{Kluson:2011tb}..
 
  This paper is organized as the following way, the action and its equivalent scalar tensor action is defined in section 2. In section 3, we review ADM formalism\cite{Arnowitt:1962hi, Gourgoulhon:2007ue}. Then we convert our scalar tensor action in ADM variable. In section 4, we perform Hamiltonian formalism for a general action where we calculate all the constraints and count the degree of freedom.  In section 5, we study the general action for some cases. Finally, we summarise our results in section 6. 
\section{Quadratic Non-local Gravity}
Let us start with constructing a generalized non-local action consisting of  invariants of Ricci scalars, Ricci tensors and Riemann Tensors. The action can be written as\cite{PhysRevD.88.123502, PhysRevD.95.043539, PhysRevD.98.084040},
\begin{eqnarray}
S=\int d^{4}x \sqrt{-g}\Big[ M_{p}^{2} R+ R \mathcal{F}_{1}(\Box) R +  R_{\mu \nu}\mathcal{F}_{2}(\Box)R^{\mu \nu} +  R_{\mu \nu \rho \sigma}\mathcal{F}_{3}(\Box)R^{\mu \nu \rho \sigma}\Big], \label{action}
\end{eqnarray}
 where 
\begin{eqnarray}
\mathcal{F}_{i}= \sum_{n=1}^{\infty} f_{i_{-n}}\Box^{-n},\qquad i= \{1,2,3 \}
\end{eqnarray}
and $ f_{i_{-n}} $ is the mass scale associated with non-local corrections.  We take $M_{p}^{2} = 1 $ for our paper.  $\Box^{-1} $ is  the inverse of a d' Alembertian operator.  Our notations is similar to that of ref.\cite{Teimouri:2018ogt}.\\
Now we divide action (\ref{action}) in four parts: $S_0,S_1,S_2,S_3 $ which are,
\begin{center}
\begin{eqnarray}
S_0 &=& \int d^{4}x \sqrt{-g}R  \label{cf},\\ 
S_1 &=& \int d^{4}x \sqrt{-g}R \mathcal{F}_{1}(\Box) R \label{s1}, \\
S_2 &=& \int d^{4}x \sqrt{-g} R_{\mu \nu}\mathcal{F}_{2}(\Box)R^{\mu \nu}  \label{s2}, \\ 
S_3 &=& \int d^{4}x \sqrt{-g} R_{\mu \nu \rho \sigma}\mathcal{F}_{3}(\Box)R^{\mu \nu \rho \sigma} \label{s3}.
\end{eqnarray}
\end{center}
Further,  if we take n=0 then action (\ref{action}) reduces to quadratic gravity action containing terms upto second order in curvature. For other positive values of $n,$ nonlocality plays a major role. In next section we write an equivalent action for this nonlocal action. 
\subsection{Equivalent Scalar tensor Action}
In order to express action (\ref{action}) in terms of an equivalent action, 
we wish to translate non-local action to scalar tensor one by defining the  
auxiliary scalar and tensor fields. Now we write the equivalent scalar 
tensor form of $S_0, S_1, S_2.$
\subsubsection{Equivalent Action  of $S_0$} For deriving equivalent scalar-tensor 
form of actions (\ref{cf}), we replace $R$ by $Q$ via a Lagrange multiplier $C$ and rewrite the action \label{s0} in the following form,
\begin{eqnarray}
S_{0}^{eqv} &=& \int d^{4}x \sqrt{-g}\Big[Q + C(R-Q)\Big]. \label{s0eq}
\end{eqnarray}
By varying action with respect to auxiliary field $C,$ we obtain $R = Q.$ Thus it is easy to check  that eq.(\ref{s0eq}) is equivalent to eq.(\ref{cf}).
\subsubsection{Equivalent Action  of $S_1$} 
To convert action (\ref{s1}) into its equivalent scalar tensor form first we put $ \mathcal{F}_{1}(\Box) = \sum_{n = 0}^{\infty}f_{1_{-n}} \Box^{-n} $, so we get
\begin{equation}
S_1 = \int d^{4}x \sqrt{-g}R \sum_{n = 0}^{\infty}f_{1_{-n}} \Box^{-n}  R.
\end{equation}
Now we replace R  by Q via Lagrange multiplier so we get,
 \begin{equation}
 S_{1}^{eqv} = \int d^{4}x \sqrt{-g} \Bigg[ Q \sum_{n=1}^{\infty}f_{1_{-n}}\frac{1}{\Box^{n}}Q + C(R-Q) \Bigg],\label{s11}
\end{equation} 
 By introducing two different set of auxiliary fields $ B_{n} $ and  $ A_{n}$, where $(n=1,2,3....\infty)$ such that $ \frac{1}{\Box} Q = A_{1} \Longrightarrow  \Box A_{1} = Q $ via Lagrange multiplier $B_{1}$ and $ \frac{1}{\Box} A_{1}=A_2 \Longrightarrow  \Box A_{2} = A_1$ via Lagrange multiplier $B_2$ so on... $ \frac{1}{\Box} A_{n-1}=A_n \Longrightarrow  \Box A_{n} = A_{n-1}$ via Lagrange multiplier $B_n $. Thus, we can rewrite the action eq.(\ref{s11}), as
\begin{equation}
\begin{split}
S_{1}^{eqv} = \int \sqrt{-g}\Bigg[ Q\Big( f_{1_{-0}} Q + \sum_{n=1}^{\infty}f_{1_{-n}}A_{n}\Big) + C\Big( R -Q\Big) + B_{1}(\Box A_{1}-Q) + \sum_{L=2}^{\infty} B_{L}(\Box A_{L} -A_{L-1}) \Bigg].
\end{split}
\end{equation}
After rearranging the above action we get,
\begin{equation}
\begin{split}
S_{1}^{eqv} = \int \sqrt{-g}\Bigg[ Q \Big(f_{1_{-0}}Q + \sum_{n=1}^{\infty} f_{1_{-n}}A_{n}\Big) + C\Big( R -Q\Big) -B_{1}Q -\sum_{L=2}^{\infty} B_{L} A_{L-1}+\sum_{L=1}^{\infty} B_{L}\Box A_{L} \Bigg]. \label{s1eq}
\end{split}
\end{equation}
\subsubsection{Equivalant Action  of $S_2$} 
Similarly for action (\ref{s2}) after substituting $ \mathcal{F}_{1}(\Box) = \sum_{n = 0}^{\infty}f_{2_{-n}} \Box^{-n}, $ we obtain,
\begin{equation}
S_2 = \int d^{4}x \sqrt{-g} R_{\mu \nu}\sum_{n = 0}^{\infty}f_{2_{-n}} \Box^{-n}R^{\mu \nu}.
\end{equation}
We replace $R^{\mu \nu}$ by $Q^{\mu\nu}$ by imposing a constraint via a Lagrangian multiplier $C _{\mu \nu}$, so we get an equivalent action,
\begin{eqnarray}
S_{2}^{eqv}=\int d^{4}x \sqrt{-g}\Big[ \sum_{n=1}^{\infty} f_{2_{-n}} Q_{\mu \nu}\frac{1}{\Box^{n}}Q^{\mu \nu} + C_{\mu \nu}(R^{\mu \nu} -Q^{\mu \nu})\Big].\label{eq}
\end{eqnarray}
Again we replace $ \frac{1}{\Box}Q^{\mu \nu} $ by $ A^{\mu \nu} $ imposing a constraint $\Box A^{\mu \nu}-Q^{\mu \nu}  $ via Lagrange multiplier $ B_{\mu \nu}$,  in next step $ \Box A^{\mu \nu}_2-A^{\mu \nu},$ via $B_{2\mu \nu},$ and so on, finally the equivalent action becomes, 
\begin{eqnarray}
S_{2}^{eqv} =\int d^{4}x \sqrt{-g}\Big[Q_{\mu \nu}(f_{2_{-0}}Q^{\mu \nu}+ \sum_{n=1}^{\infty} f_{2_{-n}}A_{n}^{\mu \nu})+C_{\mu \nu}(R^{\mu \nu} -Q^{\mu \nu})-\nonumber \\B_{1 \mu \nu}Q^{\mu \nu}-\sum_{L=2}^{\infty} B_{L \mu \nu}A_{L-1}^{\mu \nu}+\sum_{L=1}^{\infty} B_{L \mu \nu}\Box A_{L}^{\mu \nu}\Big]. \label{s2eq}
\end{eqnarray}
\section{ADM Formalism}
In order to formulate the Hamiltonian analysis of  the above theory first we review basics of 3 + 1 ADM formalism here. Suppose that (M,$ g_{\mu\nu} $) is four dimensional manifold, can be foliated by family of space like surface ($\Sigma_{t}$). In this formalism the 4-dimensional metric can be written in terms of induced metric $ h_{\mu\nu}$ on the 3-dimensional surface and normal vector $ n_{\mu}.$ $g_{\mu\nu}$ is related to induced metric $ h_{\mu\nu}$ by
\begin{equation} g_{\mu\nu}= h_{\mu\nu}- n_{\mu} n_{\nu}  \label{metric}, \end{equation}
where $ n_{\mu} $ is the time like future directed vector normal to three dimensional spacelike surface, whose norm is,
$n_{\mu}n^{\mu}=-1$\\
Any symmetric second rank tensor $T_{\mu \nu}$ in 3+1 notation can be decompose as,
\begin{eqnarray}
T_{\mu \nu}=\delta_{\mu}^{\sigma} \delta_{\nu}^{\rho}T_{\sigma\rho}=h_{\mu}^{\sigma} h_{\nu}^{\rho}T_{\sigma\rho}-n_{\mu}n^{\sigma} h_{\nu}^{\rho}T_{\sigma\rho}-h_{\mu}^{\sigma} n_{\nu}n^{\rho}T_{\sigma\rho}+n_{\mu}n^{\sigma} n_{\nu}n^{\rho}T_{\sigma\rho}\\=\tilde{T}_{\mu \nu}-n_{\mu}\mathcal{T}_{\nu}-n_{\nu}\mathcal{T}_{\mu}+n_{\mu} n_{\nu}\tau \qquad \qquad \qquad \qquad \qquad \label{tensor}
\end{eqnarray} 
where $\tilde{T}_{\mu\nu}=h_{\mu}^{\sigma} h_{\nu}^{\rho}T_{\sigma\rho}$ is purely spatial part, $ \mathcal{T}_{\mu}=n^{\sigma} h_{\nu}^{\rho}T $ is one normal projection and $ \tau=n^{\sigma}n^{\rho}T_{\sigma\rho} $ is two normal projection of tensor $ T_{\mu\nu} $.
\\
The line element in ADM formalism  is,
\begin{equation}
ds^2= -N^{2}dt^{2}+h_{ij}(dx^{i}+N^{i}dt)(dx^{j}+N^{j}dt)
\end{equation} 
Here $ N^{i}$ is shift vector and N is the lapse function, defined in terms of metric as,
\begin{equation}
N=\frac{1}{\sqrt{-g^{00}}},\qquad N^i=-\frac{g^{0i}}{\sqrt{g^{00}}}.
\end{equation}
In terms of metric variable, we have
\begin{eqnarray}
g_{00}=-N^2+h_{ij}N^{i}N^{j},\qquad g_{0i}=N_{i},\qquad g_{ij}=h_{ij}, \nonumber\\ g^{00}=\frac{1}{N^{2}},\qquad g^{0i}=\frac{N^i}{N^2}, \qquad g^{ij}=h^{ij}-\frac{N^i N^j}{N^2}.
\end{eqnarray}
In coordinate basis, 
$$ n_{0}=-N, \qquad n_{i}=0, \qquad n^{0}=\frac{1}{N}, \qquad n^{i}=\frac{N^{i}}{N}.$$
Now we can express the projection relations for the
curvature tensors in terms of ADM variables. Ricci scalar takes the form 
\begin{equation}
R= \mathcal{R}+ K^{2}-3 K_{i j} K^{ij}+ 2 h^{ij} L_{n}  K_{ij} -\frac{2}{N}\mathcal{D}_{i}\mathcal{D}^{i}N. \label{RIS}
\end{equation}
Similarly Ricci tensor takes the form
\begin{eqnarray}  {}_{\perp}R_{ij}= \mathcal{R}_{ij}+ K_{ij} K - 2K_{i k} K^{k}_{j}+  L_{n}K_{ij} - \frac{1}{N}\mathcal{D}_{i} \mathcal{D}_{j}N,\\
{}_{\perp}R_{i n}=\mathcal{D}_{k}K_{i}^{k}-\mathcal{D}_{i} K, \qquad \qquad \qquad \qquad\\
{}_{\perp}R_{n n}=  K_{ij} K^{ij}- h^{ij} L_{n}K_{ij} +\frac{1}{N} \mathcal{D}_{i}\mathcal{D}^{i}N,\label{RIT}
\end{eqnarray}  
where $ {}_{\perp}R_{ij} $ purely spatial, ${}_{\perp}R_{i n}$ one normal and $ {}_{\perp}R_{n n} $ two normal projection  of Ricci tensor, and $
\mathcal{D}_{i}$ is 3-dimensional covariant derivative 
and $ K_{ij}$ is Extrinsic curvature tensor and  $ L_{n}K_{ij}$  it's Lie derivative.
\subsection{ ADM Decomposition of Action }
In this section we decompose actions (\ref{s0eq}), (\ref{s1eq}) and (\ref{s2eq}) into ADM variables. We have already shown decomposed form of the curvature tensor in terms of ADM variables in previous section.
\subsubsection{ADM Decomposition of $S_{0}$ and $ S_{1}$}
We combine $S_{0}^{eqv}$ and $S_{1}^{eqv} $  as both contain Ricci scalar term  in a single action 
\begin{equation}
S_{01}^{eqv}  =  S_{0}^{eqv} + S_{1}^{eqv};
\end{equation} 
next we analyse the decomposition of each term on $ S_{1}^{eqv} $ and  $ S_{2}^{eqv} $.
\begin{itemize}
\item Decomposition of $C( R -Q)$:\\
$ R $ contain second order derivative of metric. However this second derivative of metric can be eliminated by performing integration by parts, so using eq.(\ref{RIS}),
\begin{equation}
 \int d^{4}x N\sqrt{h} C( R -Q)=\int d^{4}x N\sqrt{h} \Big[C(\mathcal{R}+K^{2}-2 K_{i j} K^{ij}+ 2 h^{ij} L_{n}  K_{ij} -\frac{2}{N}\mathcal{D}_{i}\mathcal{D}^{i}N-Q) \Big].
\end{equation}
After simplification,
\begin{equation}
\int d^{4}x N\sqrt{h} C( R -Q)=\int d^{4}x N\sqrt{h} \Big[C( K_{ij}K^{ij} -K^{2} + \mathcal{R} -Q) -2K \nabla_{n}C\Big].
\end{equation}
\item Decomposition of $\sum_{L=1}^{\infty} B_{L }\Box A_{L}$ becomes:
\end{itemize}
\begin{equation}
\int d^{4}x N\sqrt{h}\sum_{L=1}^{\infty} B_{L }\Box A_{L}=\int d^{4}x  N\sqrt{h}\sum_{L=1}^{\infty}\nabla _{\rho}B \nabla^{\rho} A_{L},
\end{equation}
After some mathematical calculation, we get,
\begin{equation}
  \int d^{3}x N\sqrt{h}  \Big[h^{ij} \sum_{L=1}^{\infty}  \mathcal{D}_{i}B_{L}\mathcal{D}_{j}A_{L}   -\sum_{L=1}^{\infty}  \nabla_{n}B_{L}\nabla_{n}A_{L}  \Big].
\end{equation}
By using eq.(27) and eq.(29), we obtain the final form of action (25),
\begin{eqnarray}
S_{01}^{eqv}= \int d^{3}x N\sqrt{h} \Bigg[ Q  + Q \Big( f_{1_{-0}}Q + \sum_{n=1}^{\infty}f_{1_{-n}}A_{n}\Big) + C\Big( K_{ij}K^{ij} -K^{2} + \mathcal{R} -Q\Big)\nonumber\\ -2K \nabla_{n}C  -  \sum_{L= 2}^{\infty}  B_{L} A_{L-1} - B_{1}Q + h^{ij} \sum_{L=1}^{\infty}  \mathcal{D}_{i}B_{L}\mathcal{D}_{j}A_{L} - \sum_{L=1}^{\infty}  \nabla_{n}B_{L}\nabla_{n}A_{L}\Bigg]\qquad
\end{eqnarray}
\subsubsection{ADM Decomposition of $S_{2}$}
 Similarly here while writing the  actions  eq.(\ref{s2eq}) in 3+1 formalism, we need to
decompose all the symmetric second rank tensors($R_{\mu \nu}, Q_{\mu \nu}, A_{\mu \nu}, B_{\mu \nu}, C_{\mu\nu} $) into components that are tangent and normal to the hypersurfaces using projection relations is provided in eq.(\ref{tensor}).
\begin{itemize}
\item Decomposition of $f_{1_{-0}}Q_{\mu \nu}Q^{\mu \nu}$:
\begin{eqnarray}
f_{2_{-0}}Q_{\mu \nu}Q^{\mu \nu}=f_{2_{-0}}(\tilde{Q}_{ij}\tilde{Q}^{ij}-2\mathcal{Q}_{i}\mathcal{Q}^{i} +\theta^{2}),\label{si}
\end{eqnarray}
where
$$ \tilde{Q}_{ij}=h^{\mu}_{i}h^{\nu}_{ j }Q_{\mu \nu}, \qquad \mathcal{Q}_{i}=h_{i}^{\mu}n^{\nu}Q_{\mu \nu}, \qquad \theta=n^{\mu}n^{\nu}Q_{\mu \nu}. $$\\ 
$ \tilde{Q}_{ij} $ is purely spatial part and $\mathcal{Q}_{i}$  and $ \theta $ are one normal, two  normal part of $ Q_{\mu\nu} $ respectively.
\item Decomposition of $\sum_{L=1}^{\infty} f_{2_{-n}}A_{n}^{\mu \nu}Q_{\mu \nu} $:
\begin{eqnarray}
\sum_{n=1}^{\infty} f_{2_{-n}}A_{n}^{\mu \nu}Q_{\mu \nu}=
 \sum_{n=1}^{\infty} f_{2_{-n}}\Big(\tilde{Q}_{ij}\tilde{A}^{ij}_{n}-2\mathcal{Q}_{i}\mathcal{A}^{i}_{n}+\theta \Lambda_{n}\Big),
\end{eqnarray} 
where,
$$ \tilde{A}^{ij}_{n}=h^{i}_{\mu}h^{j}_{\nu}A^{\mu\nu}_{n},\qquad \mathcal{A}^{i}_{n}=h^{i}_{\mu}n_{\nu}A^{\mu\nu}_{n}, \qquad \Lambda_{n}= n_{\mu}n_{\nu}A^{\mu \nu}_{n}.$$
$\tilde{A}^{ij}_{n}$,$\mathcal{A}^{i}_{n} $ and $\Lambda_{n}$ are spatial, one normal and two normal decomposition of $A_{n}^{\mu \nu}$ respectively. 
\item Decomposition of $B_{1 \mu \nu}Q^{\mu \nu}$:
\begin{eqnarray}
B_{1 \mu \nu}Q^{\mu \nu}=\Big(\tilde{B}_{1ij}\tilde{Q}^{ij}-2\mathcal{B}_{1i}\mathcal{Q}^{i}+\Psi_{1}\theta\Big).
\end{eqnarray}
\item Decomposition of $ \sum_{L=2}^{\infty} B_{L \mu \nu}A_{L-1}^{\mu \nu} $ is,
\begin{eqnarray}
\sum_{L=2}^{\infty} B_{L \mu \nu}A_{L-1}^{\mu \nu}=\sum_{L=2}^{\infty}(\tilde{B}_{Lij}\tilde{A}^{ij}_{L-1}-2\mathcal{B}_{Li}\mathcal{A}^{i}_{L-1}+\Psi_L \Lambda_{L-1}),
\end{eqnarray}
where,
$$ \tilde{B}_{Lij}=h_{i}^{\mu}h_{j}^{\nu}B_{L\mu \nu},\qquad \mathcal{B}_{Li}=h_{i}^{\mu} n^{\nu}A_{L \mu \nu}, \qquad \Psi_{L}= n^{\mu}n^{\nu}B_{L\mu \nu}.$$
\item{Decomposition of $C_{\mu \nu}(R^{\mu \nu} -Q^{\mu \nu})$:} \\
Here note that $ R^{\mu\nu} $ contain second order derivative of metric. However this second derivative of metric can be eliminated by performing integration by parts.
The part of the action we are interested in here, is
\begin{eqnarray}
S_{H} = \int d^{4}x \sqrt{-g}C^{\mu \nu}(R_{\mu \nu} -Q_{\mu \nu}),\label{a1}
\end{eqnarray}
The 3+1 decomposition of $C^{\mu\nu}(Q_{\mu\nu}-R_{\mu\nu})$ is given by,
\begin{eqnarray}
C^{\mu \nu}(R_{\mu \nu} -Q_{\mu \nu})=\tilde{C}^{ij}({}_{\perp}R_{ij}-\tilde{Q}_{ij})-2 \mathcal{C}^{i}({}_{\perp}R_{i n}-\mathcal{Q}_{i})+{\Omega}({}_{\perp}R_{n n}-\theta),
\end{eqnarray}  
where 
$$ \tilde{C}^{ij}=h_{\mu}^{i}h_{\nu}^{ j }C^{\mu \nu},\qquad \mathcal{C}^{i}=h^{i}_{\mu}n_{\nu}C^{\mu \nu}, \qquad \Omega= n_{\mu}n_{\nu}C^{\mu \nu}.$$ 
Then by using eq.(\ref{RIT}), $S_{H}$ takes the form
\begin{eqnarray}
S_{H}=\int d^{4}x N\sqrt{h}[C^{ij}(\mathcal{R}_{ij}+ K_{ij} K - 2K_{i s} K^{s}_{j}+  L_{n}  K_{ij} - \frac{1}{N} D_{i} D_{j}N-\tilde{Q}_{ij})\nonumber \\ -2C^{i}(D_{s}K_{i}^{s}-D_{i} K-\mathcal{Q}_{i})+{\Omega}( K_{s t} K^{s t}- h^{s t} L_{n}  K_{s t} +\frac{1}{N} D_{s} D^{s}N-\theta)].
\end{eqnarray}
After eliminating the higher derivative term we can write  the action ($ S_H $) as, (see appendix A for details)
\begin{eqnarray}
S_{H}=\int d^{4}x N\sqrt{h}[C^{ij}(\mathcal{R}_{ij}-\tilde{Q}_{ij})-2C^{i}(D_{s}K_{i}^{s}-D_{i} K-\mathcal{Q}_{i})\nonumber \\ +{\Omega}(- K_{s t} K^{s t}+K^{2} -\theta)-(\nabla_{\bold{n}}C^{ij})K_{ij}+(\nabla_{\bold{n}}\Omega)K],\qquad \qquad
\end{eqnarray}
\item Decomposition of $\sum_{L=1}^{\infty} B_{L \mu \nu}\Box A_{L}^{\mu \nu}$: \\ 
In this case by integrating by parts, action reduces to
\begin{eqnarray}
S_{H1}=\int d^{4}x \sqrt{-g}\sum_{L=1}^{\infty} B_{L \mu \nu}\Box A_{L}^{\mu \nu}=\int d^{4}x \sqrt{-g}\sum_{L=1}^{\infty}\nabla _{\rho}B_{L \mu \nu} \nabla^{\rho} A_{L}^{\mu \nu} .\
\end{eqnarray} 
The full analysis of decomposition of term $ \nabla _{\rho}B_{L \mu \nu} $ is given in the appendix B. The final decomposed expression of  $ \nabla _{\rho}B_{L \mu \nu} $ is,
\begin{eqnarray}
 \nabla _{\rho}B_{L \mu \nu} =D _{\rho}\tilde{B}_{L \mu \nu}-K_{\rho\mu}\mathcal{B}_{L \nu}-\mathcal{B}_{L \mu} K_{\rho\nu}-n _{\rho}\Big[\nabla_{n}\tilde{B}_{L \mu \nu}-a_{\mu}\mathcal{B}_{L \nu}-a_{\nu}\mathcal{B}_{L \mu}\Big]-\nonumber \\ n_{\mu} \Big[D _{\rho}\mathcal{B}_{L \nu}-\Psi K_{\rho\nu}\Big] -n_{\nu}\Big[D _{\rho}\mathcal{B}_{L \mu} -\Psi K_{\rho\mu}\Big]-n_{\mu}n_{\rho}\Big(a_{\nu}-\nabla_{n}\mathcal{B}_{L \nu}\Big)\nonumber \\-n_{\nu}n_{\rho}\Big(a_{\mu}-\nabla_{n}\mathcal{B}_{L \nu}\Big)+n_{\mu}n_{\nu}\nabla_{\rho}\Psi.
\end{eqnarray}
\end{itemize}
A similar expression can be obtained for $ \nabla _{\rho}A^{\mu \nu}_{L}. $ Hence the action $S_{H1}$ becomes,
\begin{eqnarray}
S_{H1}=\int d^{4}x \sqrt{-g}\Big[\sum_{L=1}^{\infty}(D _{\rho}\tilde{B}_{L \mu \nu}-K_{\rho\mu}\mathcal{B}_{L \nu}-\mathcal{B}_{L \mu} K_{\rho\nu})(D ^{\rho}\tilde{A}^{\mu \nu}_{L}-K^{\rho\mu}\mathcal{A}^{\nu}_{L}-\mathcal{A}^{\mu}_{L} K^{\rho\nu})+\nonumber\\(\nabla_{n}\tilde{B}_{L \mu \nu}+a_{\mu}\mathcal{B}_{L \nu}+a_{\nu}\mathcal{B}_{L \mu})(\nabla_{n}\tilde{A}^{\mu \nu}_{L}+a^{\mu}\mathcal{A}^{\nu}_{L}+a^{\nu}\mathcal{A}^{\mu}_{L}) +4\Big( D_{\rho}\mathcal{B}_{L \nu}- \Psi K_{\rho\nu}\Big) \nonumber\\\Big(D ^{\rho}\mathcal{A}^{\nu}_{L}-\Lambda K^{\rho\nu}\Big)-4\Big(a_{\nu}-\nabla_{n}\mathcal{B}_{L \nu}\Big)\Big(a^{\nu}-\nabla^{n}\mathcal{A}^{\nu}_{L} \Big)+D_{\rho}\Psi_{L} D^{\rho}\Lambda_{L}-\nabla_{n}\Psi_{L} \nabla_{n}\Lambda_{L} \Big].
\end{eqnarray}
The complete decomposed form of action $S_2^{eqv}$ becomes,
\begin{eqnarray}
S_2^{eqv} = \int d^{4}x N\sqrt{h}\Bigg[f_{2_{-0}}\Big(\tilde{Q}_{ij}\tilde{Q}^{ij}-2\mathcal{Q}_{i}\mathcal{Q}^{i} +\theta^{2}\Big)+\sum_{n=1}^{\infty} f_{2_{-n}}\Big(\tilde{Q}_{ij}\tilde{A}^{ij}_{n}-2\mathcal{Q}_{i}\mathcal{A}^{i}_{n}+\nonumber\\\theta \Lambda_{n}\Big)+C^{ij}\Big(\mathcal{R}_{ij}-\tilde{Q}_{ij}\Big)-2C^{i}\Big(D_{s}K_{i}^{s}-D_{i} K-\mathcal{Q}_{i}\Big) +{\Omega}\Big(- K_{s t} K^{s t}+K^{2}-\theta\Big) \nonumber \\-(\nabla_{\bold{n}}C^{ij})K_{ij}+(\nabla_{\bold{n}}\Omega)K +\tilde{B}_{1ij}\tilde{Q}^{ij}-2\mathcal{B}_{1i}\mathcal{Q}^{i}+\Psi_{1}\theta +\sum_{L=2}^{\infty}\Big(\tilde{B}_{lij}\tilde{A}^{ij}_{L-1}-2\mathcal{B}_{il}\mathcal{A}^{i}_{L-1}\nonumber \\+\Psi_L \Lambda_{L-1}\Big)+\sum_{L=1}^{\infty}\Bigg\{(D _{k}\tilde{B}_{L i j}-K_{ki}\mathcal{B}_{L j}-\mathcal{B}_{L i} K_{kj})(D ^{k}\tilde{A}^{i j}_{L}-K^{ki}\mathcal{A}^{j}_{L}-\mathcal{A}^{i}_{L} K^{kj})+\nonumber\\(\nabla_{n}\tilde{B}_{L i j}+a_{i}\mathcal{B}_{L j}+a_{j}\mathcal{B}_{L i})(\nabla_{n}\tilde{A}^{i j}_{L}+a^{i}\mathcal{A}^{j}_{L}+a^{j}\mathcal{A}^{i}_{L}) +4\Big( D_{k}\mathcal{B}_{L j}- \Psi K_{kj}\Big)\Big(D ^{k}\mathcal{A}^{j}_{L}\nonumber\\-\Lambda K^{kj}\Big)-4\Big(a_{j}-\nabla_{n}\mathcal{B}_{L j}\Big)\Big(a^{j}-\nabla^{n}\mathcal{A}^{j}_{L} \Big)+D_{k}\Psi_{L} D^{k}\Lambda_{L}-\nabla_{n}\Psi_{L} \nabla_{n}\Lambda_{L} \Bigg\}\Bigg] \label{s2eqv}
\end{eqnarray}
Note that the above action is a function of  $(h_{ij}, N, N^i,
\tilde{Q}_{ij},\mathcal{Q}_{i},\theta, \tilde{A}_{ij},\mathcal{A}_{i},\Lambda,\tilde{B}_{ij},\mathcal{B}_{i},\Psi,C^{ij},C^{i},\Omega$) and their time and space derivatives.
\section{Hamitonian Analysis}
\subsection{Hamitonian for $S_{01}$}
Having expressed $S_{01}$ in terms of $(h_{ij}, N, N^{i}, Q, C, A_{L}, B_{L})$ and their space and time derivatives, we are now ready to carry out the Hamiltonian analysis of this action. Here
we introduce canonical conjugate momenta corresponding to each variable. Canonical momenta with respect to set $(N, N^i, h_{ij}, Q, C, A_{L}, B_{L})$ are 
\begin{eqnarray*}
\Pi_{N} \approx 0, \qquad\qquad
\Pi_{i} \approx 0, \qquad 
\Pi^{ij} = \sqrt{h} C( \mathcal{K}^{ij} -h^{ij}\mathcal{K} ) - \sqrt{h} h^{ij}\nabla_{n}C, \qquad \qquad \qquad \qquad   \\
P_{Q} \approx  0 , \qquad \qquad
P_{C} = -2\sqrt{h} \mathcal{K}, \qquad
P_{A_{L}} = -\sqrt{h} \sum_{L= 1}^{\infty} \nabla_{n} B_{L}, \qquad  \qquad
P_{B_{L}} = -\sqrt{h} \sum_{L= 1}^{\infty} \nabla_{n} A_{L}.
\end{eqnarray*}
Here we used the $\approx$ sign for showing the primary constraint and they valid only on constraint surface $\Gamma$. 
Hence the primary constraints are summarised as,
\begin{eqnarray*}
\Pi_{N}\approx 0 , \Pi_{i}\approx 0 , P_{Q}\approx 0.
\end{eqnarray*}
 It can be observed that the Lagrangian density $ \mathcal{L} $ does not contain the time derivative of  $ N, N^{i} $ and $Q$. Hence they are no longer dynamical variables. Hamiltonian density is given by,
\begin{eqnarray}
\mathcal{H}=\Pi^{ij}\dot{h}_{ij}+p^C \dot{C} +\sum_{L = 1}^{\infty}( p^{B_L}\dot{B}_{L} + p^{A_L} \dot{A}_{L})-\mathcal{L}.
\end{eqnarray}
By splitting $\mathcal{H}$ along $N$ and $N_i$ direction, we obtain,
\begin{equation}
\mathcal{H}= N\mathcal{H}_{N}  + N^{i} \mathcal{H}_{i,} 
\end{equation}
with
\begin{equation}
\begin{split}
\mathcal{H}_{N} & = \frac{1}{\sqrt{h} C} \Pi^{ij} h_{ik} h_{jl} \Pi^{kl} -\frac{1}{3\sqrt{h} C} \Pi^{2} -\frac{1}{3\sqrt{h}} \Pi P_{C} + \frac{1}{6 \sqrt{h}} C P_{C}^{2} -\sqrt{h} C \mathcal{R}+ \sqrt{h}C Q  \\& - \sqrt{h} Q \Big( 1 + f_{1_{-0}}Q + \sum_{L=1}^{\infty}f_{1_{-n}}A_{n}\Big) +\sqrt{h}\sum_{L=2}^{\infty} A_{L-1}B_{L} +\sqrt{h}B_{1}Q \\&  -\sum_{L = 1}^{\infty} \frac{P_{A_{L}}P_{B_{L}}}{\sqrt{h}} -\sqrt{h} h^{ij} \sum_{L = 1}^{\infty} \Big( \mathcal{D}_{i}B_{L} \mathcal{D}_{j}A_{L}\Big),  \label{hnso1}
\end{split}
\end{equation}
and
\begin{equation}
\begin{split}
\mathcal{H}_{i}  = -2 h_{ik}\mathcal{D}_{j}\Pi^{kj} + \sum_{L = 1}^{\infty} \Big(P_{A_{L}}\mathcal{D}_{i}A_{L} + P_{B_{L}}\mathcal{D}_{i}B_{L} \Big) + P_{C}\mathcal{D}_{i}C.    \label{hiso1}
\end{split}
\end{equation} 
Now the Hamiltonian density is given by,
\begin{equation}
\mathcal{H}_{tot} =  N\mathcal{H}_{N} + N^{i}\mathcal{H}_{i} +\lambda^{i}\Pi_{i}+\lambda^{N}\Pi_{N} + \lambda^{Q}P_{Q},
\end{equation}where $ \lambda^{i},\lambda^{N} $ are Lagrange multipliers.
and total Hamiltonian  becomes,
\begin{equation}
H_{total} = \int d^{3}x \Big( N \mathcal{H}_{N}  + N^{i} \mathcal{H}_{i} +\lambda^{i}\Pi_{i}+\lambda^{N}\Pi_{N} + \lambda^{Q}P_{Q} \Big),
\end{equation}

\subsubsection{Classification of Constraints}
Here we classify the constraints according to their nature. Our primary 
constraints are $ \Pi_{N}\approx 0, \Pi_{i}\approx 0 , P_{Q}\approx 0$. 

Next the secondary constraints are determined by:
$H_N \approx 0$, $ H_i \approx 0$,
On the constraint space $ \Gamma $, the time evolutions of $H_N$ and $H_i$ vanishes weakly as
\begin{equation}
\dot{\mathcal{H_N}}=\{\mathcal{H}_N,H_{tot}\}\approx0,
\end{equation} 
and
\begin{equation}
\dot{\mathcal{H}}_i=\{\mathcal{H}_i,H_{tot}\}\approx0.
\end{equation} 
These condition  turn  fix the Lagrange multipliers $ \lambda^{i}$, $ \lambda^{N}$and there will be no tertiary constraints.
Now we have identified the primary and secondary constraints. Then we categorise
them into first and second-class constraints.  We evaluate Poisson bracket between primary constraint with total Hamiltonian. 
\begin{eqnarray}
G_{Q} = \partial_{t}P_{Q} =\{ P_{Q}, \mathcal{H}_{tot}\} = N\sqrt{h}\Big\{-C +(1 + 2f_{1_{-0}}Q+\sum_{n= 1}^{\infty} f_{1_{-n}}A_{n}) - B_{1}\Big\}.
\end{eqnarray}
Then time evolution of $G_{Q}$ become,
\begin{eqnarray}
\dot{G}_{Q} = \{ G_{Q},\mathcal{H}_{tot}\}=
 N \Bigg[ \frac{N}{3}(\Pi -CP_{C}) + N \sum_{n = 1}^{\infty} f_{1_{-n}}P_{B_{n}} 
 + P_{A_{1}}  +\\ \nonumber \sqrt{h}N^{i} \Big\{ \sum_{L=1}^{\infty} f_{1_{-L}}\mathcal{D}_{i}A_{L} + f_{1_{-1}}\mathcal{D}_{i}B_{1} \Big\}  + \sqrt{h}f_{1_{-0}}\lambda^{Q} \Bigg].
\end{eqnarray}
From the equation of motion for $ B_L $, we derive
\begin{equation}
A_1=\Box Q;\qquad \qquad A_L=\Box A_{L-1}=\Box^{L}Q,
\end{equation}
for $ L\geq2. $ Therefore we conclude that the $ B_L(L\geq2) $ are Lagrange multipliers and further we obtain other primary constraints which are
\begin{eqnarray}
Z_{1} &=& A_{1} - \Box Q = 0, \\
Z_{L} &=& A_{L} - \Box A_{L-1} = 0.
\end{eqnarray}
 Moreover $  A_{1} - \Box Q \approx 0  $ and $ A_{L} - \Box A_{L-1} \approx 0 $  are primary constraints since the Poisson brackets of these quantities with total Hamiltonian  vanish weakly by using their equation of motion. 
Whereas $P_Q$ is a second class constraint since,
\begin{equation}
\{P_Q,G_Q\}=\sqrt{h}Q\neq 0.
\end{equation}
All other Poisson brackets are obtained as,
\begin{eqnarray}
\{ \Pi_{N}, \Pi_{i}\}= \{ \Pi_{N}, \Pi_{N}\} = \{ \Pi_{N}, \mathcal{H}_{N}\}=\{ \Pi_{N},\mathcal{H}_{i}\} \approx 0 ,\nonumber \\
 \{ \Pi_{i}, \Pi_{i}\} = \{ \Pi_{i},\mathcal{H}_{N}\} = \{ \Pi_{i},\mathcal{H}_{i}\} \approx 0 ,\nonumber \\
 \{ P_{Q}, P_{Q}\} = \{ P_{Q}, \Pi_{N}\} = \{ P_{Q}, \Pi_{i}\} \approx 0.
 \end{eqnarray}
In summary, we have four first class constraints $(\Pi_{N}, \Pi_{i}, \mathcal{H}_{N}, \mathcal{H}_{i})$ and two second class constraints $P_Q $ and $ G_Q.$
\subsubsection{Degrees of Freedom}
We use the general formula to count the number of the physical degrees of freedom,
\begin{equation}
 N=\frac{1}{2}(2A-B-2C),\end{equation}
Where,\\ 
A = number of phase space variables,\\
B = number of second-class constraints,\\
C = number of first-class constraints.\\

For our nonlocal model, we have
\begin{eqnarray}
 A= \Bigg\{(h_{ij},\Pi_{ij}),(N,\Pi_{N}),N^{i},\Pi_{i} ,(Q,P_{Q}),(C,P_{C}),
 \underbrace{(B_{L},P_{B_{L}})}_\text{$\infty$} \Bigg\},
\end{eqnarray}
where L runs from 1 to $\infty$.
This amounts to
\begin{eqnarray}
2A &=& 2 \times ( 6 + 1 + 3 + 1+ 1 + \infty \nonumber \\
&=& 24 + \infty
\end{eqnarray}
This is expected as we have infinite number of derivative terms in the action. 
\subsection{Hamiltonian Analysis of $S_2$}
In this section, we focus on the Hamiltonian analysis for the
action $S_2  $ in  eq.(\ref{s2eqv}). First we write canonical momenta of each of 
variable appearing in the action as
\begin{eqnarray}
\Pi^{ij}_{\tilde{Q}}&=&\frac{\partial L}{\partial \dot{\tilde{Q}}_{ij}}\approx0, \qquad
\Pi^{i}_{\tilde{Q}}=\frac{\partial L}{\partial \dot{\mathcal{Q}}_{i}}\approx0, \qquad
\Pi_{\theta}=\frac{\partial L}{\partial \dot{\theta}}\approx 0 
 \nonumber \\ 
\Pi^{ij}_{\tilde{A}_{L}}&=&\frac{\partial L}{\partial \dot{\tilde{A}}_{Lij}}=
\sqrt{h}(\nabla_{n}\tilde{B}_{L}^{ i j}+a^{i}\mathcal{B}_{L}^{j}+a^{j}\mathcal{B}{L}^{i}),\qquad\Pi_{N}\approx 0,\qquad\Pi_{i}\approx 0,
 \nonumber \\
\Pi^{i}_{\mathcal{A}_{L}}&=&\frac{\partial L}{\partial \dot{\mathcal{A}}_{Li}}=4\sqrt{h}(a^{j}-\nabla_{n}\mathcal{B}_{L}^{j}),\qquad
\Pi_{\Lambda_{L}}=\frac{\partial L}{\partial \dot{\Lambda_{L}}}=-\sqrt{h}\nabla_{n}\Psi_{L},
 \nonumber \\ 
\Pi^{ij}_{C}&=&\frac{\partial L}{\partial \dot{C}_{ij}}=-\sqrt{h} K^{ij}, \qquad
\Pi^{i}_{C}=\frac{\partial L}{\partial \dot{C}_{i}}=0,\qquad
\Pi_{\Omega}=\frac{\partial L}{\partial \dot{\Omega}}=\sqrt{h}K,
 \nonumber \\ 
\Pi^{ij}_{\tilde{B}_{L}}&=&\frac{\partial L}{\partial \dot{\tilde{B}}_{ij}}=\sqrt{h}(\nabla_{n}\tilde{A}^{ i j}_L+a^{i}\mathcal{A}^{j}_{L}+a^{j}\mathcal{A}_{L}^{i}),
 \nonumber \\ 
\Pi^{i}_{\mathcal{B}_{L}}&=&\frac{\partial L}{\partial \dot{\mathcal{B}}_{Li}}=4\sqrt{h}(a^{j}-\nabla_{n}\mathcal{A}_{L}^{j}),\qquad
\Pi_{\Psi_L}=\frac{\partial L}{\partial \dot{\Psi_L}}=-\sqrt{h}\nabla_{n}\Lambda_{L},\nonumber
\end{eqnarray}
\begin{eqnarray}
\Pi_{h}^{ij}= \frac{\partial \mathcal{L}}{\partial \dot{h}_{ij}}= \sqrt{h}\Bigg[ \Omega (K h^{ij} -K^{ij}) -\frac{1}{2}(\nabla_{n} C^{ij} -h^{ij}\nabla_{n}\Omega) + \sum_{L=1}^{\infty} \Big[ 2K^{ij}\mathcal{A}_{Lk}\mathcal{B}^{k}_{L}+2K^{ik} \mathcal{A}_{L}^{j}\mathcal{B}_{Lk}\nonumber\\- h^{i s}\mathcal{D}_{s} \tilde{A}_{L}^{kj}\mathcal{B}_{k} - h^{i s}\mathcal{D}_{s} \tilde{B}_{L}^{kj}\mathcal{A}_{k}  - 2  \mathcal{D}^{i}\mathcal{A}_{L}^{j}\Psi_{L} +2\Lambda_{L} K^{i j} \Psi_{L} - 2  \mathcal{D}^{i}\mathcal{B}_{L}^{j}\Lambda_{L} +4\Psi_L K^{i j} \Lambda_{L}\Big],\label{mom}
\end{eqnarray}
where ($\approx$) is weak equality, used for identification of primary constraint. From the above expressions we conclude that $\Pi^{ij}_{\tilde{Q}},\Pi^{i}_{\tilde{Q}},\Pi_{\theta},\Pi^{i}_{C}$ are primary constraints. The primary constraints vanish only on constraint space but not everywhere. Hamiltonian density is given by,
\begin{equation} 
\begin{aligned}
\mathcal{H} &= \Pi^{ij}\dot{h}_{ij} + \sum_{L=1}^{\infty}\Big[\Pi^{ij}_{\tilde{A}_{L}}\dot{\tilde{A}}_{Lij} + \Pi^{ij}_{\tilde{B}_{L}}\dot{\tilde{B}}_{Lij} + \Pi^{i}_{\mathcal{A}_{L}}\dot{\mathcal{A}}_{Li} + \Pi^{i}_{\mathcal{B}_{L}}\dot{\mathcal{B}}_{Li} \\& + \Pi_{\Lambda_n}\dot{\Lambda}_{L} + \Pi_{\Psi}\dot{\Psi}_L\Big]+ \Pi^{ij}_{\tilde{C}}\dot{C}_{ij} + \Pi_{\Omega}\dot{\Omega}  -\mathcal{L},
\end{aligned}
\end{equation}
Using the relations eq.(\ref{mom}), we can write Hamiltonian density $ (\mathcal{H})$  as,
\begin{equation}
\mathcal{H} = N \mathcal{H}_{N} + N^{k} \mathcal{H}_k,
\end{equation}
where $ \mathcal{H}_{N} $ and $ \mathcal{H}_{k}$ are given as follows
\begin{equation}
\begin{aligned}
\mathcal{H}_{N} & = \sqrt{h} \Big\{ -f_{0} ( \tilde{Q}_{ij} \tilde{Q}^{ij} + 2 \mathcal{Q}_{i}\mathcal{Q}^{i} -\theta^{2})-\sum_{n=1}^{\infty} f_{2_{-n}}\Big(\tilde{Q}_{ij}\tilde{A}^{ij}_{n}-2\mathcal{Q}_{i}\mathcal{A}^{i}_{n}+\theta \Lambda_{n}\Big) -C^{ij}\Big(\mathcal{R}_{ij} -\tilde{Q}_{ij}\Big) \\&
-\Omega ( \Pi_{\Omega}^{2}- h_{sp}h_{tp}\Pi^{pq}_{\tilde{C}} \Pi^{st}_{\tilde{C}}-\theta)  + \tilde{B}_{1ij}\tilde{Q}^{ij} -2\mathcal{B}_{1i}\mathcal{Q}^{i} + \Psi_1 \theta +\sum_{L = 2}^{\infty} 
\Big(B_{Lij}A_{L-1}^{ij} -2B_{Li}A_{L-1}^{i} + \Lambda_{L-1}\Psi_{L} \Big)\Big\} 
\\ &
 + \frac{\Pi^{ij}_{h}}{\sqrt{h} \Omega} ( h_{ki}h_{lj}\Pi_{C}^{kl} + h_{ij} \Pi_{\Omega})
 + \Pi^{ij}_{\tilde{A}_L}\Big\{ \frac{1}{\sqrt{h}} h_{ki}h_{lj}\Pi_{\tilde{B_{L}}}^{kl} -a_i \mathcal{A}_{L_{j}}  -a_j \mathcal{A}_{L_{i}} \Big\} 
 + \Pi^{ij}_{\tilde{B}_L}\Big\{ \frac{1}{\sqrt{h}} h_{ki}h_{lj}\Pi_{\tilde{A_{L}}}^{kl} \\&  -a_i \mathcal{B}_{Lj}  -a_j \mathcal{B}_{Li} \Big\} 
-\frac{\Pi^{i}_{\mathcal{A}_L}}{4} \Big\{ \frac{1}{\sqrt{h}} h_{ki}\Pi_{\mathcal{B}_{L}}^{k} -a_i \Big\} -\frac{\Pi^{i}_{\mathcal{B}_L}}{4} \Big\{ \frac{1}{\sqrt{h}} h_{ki}\Pi_{\mathcal{A}_{L}}^{k} -a_i \Big\} + 2 \frac{1}{\sqrt{h}} \Pi_{\Psi_{L}}\Pi_{\Lambda_{L}}
-\\ &
\sum_{L=1}^{\infty}\Big[ \sqrt{h}\Big\{ \mathcal{D}_{s}\tilde{B}_{Lij} + \frac{1}{\sqrt{h}} h_{s k}h_{il} \Pi^{kl}_{C} \mathcal{B}_{Lj} + \frac{1}{\sqrt{h}} h_{s k}h_{jl} \Pi^{kl}_{C}  \mathcal{B}_{Li} \Big\}
\Big\{ \mathcal{D}^{s}\tilde{A}_{L}^{ij} + \frac{1}{\sqrt{h}}\Pi^{s i}_{C}\mathcal{A}^{j}_{L} + \frac{1}{\sqrt{h}} \mathcal{A}^{i}_{L}\Pi^{s j}_{C}\Big\} 
\\&
 + \frac{1}{\sqrt{h}} h_{ki}h_{lj}\Pi_{\mathcal{B_{L}}}^{kl}\Pi^{ij}_{\mathcal{A}_{L}}  + 4 \sqrt{h}\Big\{\mathcal{D}_{s} \mathcal{B}_{Lj} + \frac{1}{\sqrt{h}}\Psi h_{s k}h_{jl} \Pi^{kl}_{C}\Big\} 
\Big\{\mathcal{D}^{s}\mathcal{A}_{L}^{j} + \frac{1}{\sqrt{h}} \Lambda \Pi^{s j}_{C} \Big\}
\\& 
-\frac{1}{4 \sqrt{h}} \Pi^{i}_{\mathcal{A}_L} h_{ik} \Pi^{k}_{\mathcal{B}_L} -\sqrt{h} \mathcal{D}_{s} \Psi_{L} \mathcal{D}^{s} \Lambda_{L} + \frac{1}{\sqrt{h}} \Pi_{\Psi_L}  \Pi_{\Lambda_L}\Big],
\end{aligned}
\end{equation}
and
\begin{equation}
\begin{aligned}
\mathcal{H}_{k}  &= \Bigg[ \Pi^{ij}_{\tilde{A}_L} \mathcal{D}_{k}\tilde{B}_{L_{ij}}  + \Pi^{ij}_{\tilde{B}_L} \mathcal{D}_{k}\tilde{A}_{L_{ij}}  -\frac{1}{4}\mathcal{D}_{k}\mathcal{B}_{L_i} \Pi^{i}_{\mathcal{A}_L}  -\frac{1}{4}\mathcal{D}_{k}\mathcal{A}_{L_i} \Pi^{i}_{\mathcal{B}_L} + \Pi_{\Psi} \mathcal{D}_{k} \Psi_{L}  \\&+ \Pi_{\Lambda} \mathcal{D}_{k} \Lambda_{L}   + \Pi_{C_{ij}} \mathcal{D}_{k} C^{ij} - \Pi_{\Omega} \mathcal{D}_{k}\Omega  \Bigg].
\end{aligned}
\end{equation}
Then the total Hamiltonian density is given by 
\begin{equation}
\mathcal{H}_{tot} = N \mathcal{H}_{N} + N^{k} \mathcal{H}_k+\lambda^{N}\Pi_{N} +\lambda^{i}\Pi_{i}+\lambda_{\tilde{Q}_{ij}}\Pi^{ij}+\lambda_{\mathcal{Q}i}\Pi^{i}_{\mathcal{Q}_{i}}+
\lambda_{\theta}\Pi_{\theta}+\lambda_{C_{i}}\Pi^{C}_{i},
\end{equation}
where $\lambda_{N},\lambda_{i},\lambda_{\tilde{Q}_{ij}}, \lambda_{\mathcal{Q}i}, \lambda_{\theta}$ and $\lambda_{C_{i}}$ are the Lagrange multipliers associated with primary constraints and total Hamiltonian is
\begin{eqnarray}
H=\int d^3 x( N \mathcal{H}_{n} + N^{k} \mathcal{H}_k+\lambda^{N}\Pi_{N} +\lambda^{i}\Pi_{i}+\lambda_{\tilde{Q}_{ij}}\Pi^{ij}+\lambda_{\mathcal{Q}i}\Pi^{i}_{\mathcal{Q}_{i}}+
\lambda_{\theta}\Pi_{\theta}+\lambda_{C_{i}}\Pi^{C}_{i})
\end{eqnarray}
\subsubsection{Classification of Constraint}
In previous section we found conjugate momenta corresponding to their canonical variable. Now we need to check the consistency of primary constraints.
\begin{eqnarray}
\dot{\Pi_{N}}=\{\Pi_{N},\mathcal{H}_{tot}\}=H_N,~~ \qquad \dot{\Pi_{i}}=\{\Pi_{i},\mathcal{H}_{tot}\}=H_i,
\end{eqnarray}
are Hamiltonian field equations and enforce they are zero at the constraint surface so $H_N\approx 0$,$H_i\approx 0$, where $H_N$ as the Hamiltonian constraint, and $H_i$ as diffeomorphism constraint. They fulfil the criteria that of secondary constraints.\\
We define $G^{ij}_{\tilde{Q}}$ corresponding $\tilde{Q}_{ij}$ as,
\begin{equation}
G^{ij}_{\tilde{Q}}=\dot{\Pi^{ij}}_{\tilde{Q}}=\{\Pi^{ij}_{\tilde{Q}},\mathcal{H}_{tot}\}=N\sqrt{h}\Big\{2 f_{2_{-0}}Q^{ij}+\sum_{n=1}^{\infty}( f_{2_{-n}}\tilde{A}^{ij}_{n})-\tilde{C}^{ij}- \tilde{B}^{ij}\Big\}\approx 0,
\end{equation}
which acts as a secondary constraint corresponding to primary constraint $\tilde{Q}_{ij}\approx 0$ and further
\begin{equation}
\{\Pi^{ij}_{\tilde{Q}},G^{ij}_{\tilde{Q}}\}\neq0 ,
\end{equation}
implies that $\Pi^{ij}_{\tilde{Q}}, G^{ij}_{\tilde{Q}}$ are second class constraints.\\
However the evolution of constraint $G^{ij}_{\tilde{Q}}$ is
\begin{equation}
\dot{G^{ij}}_{\tilde{Q}}=\{G^{ij}_{\tilde{Q}},\mathcal{H}_{tot}\}\approx 0,
\end{equation}
which in turn implies,
\begin{eqnarray}
=N\sqrt{h}\Bigg[N\Big\{-2\sqrt{h}\Pi^{ij}_{C}-\frac{\Pi^{ij}_{h}}{\sqrt{h} \Omega} +\frac{4}{\sqrt{h}} \Pi^{ki}_{C}(\mathcal{A}^{j}\mathcal{B}_{k}+\mathcal{A}_{k}\mathcal{B}^{j})+2\mathcal{B}^{k}\mathcal{D}^{j}\mathcal{A}^{i}_{k}+2\mathcal{A}^{k}\mathcal{D}^{j}\mathcal{B}^{i}_{k} \nonumber\\ +4\mathcal{D}^{i}\mathcal{A}^{j}(\Psi +\Lambda )+\frac{2}{\sqrt{h}}\Psi \Lambda \Pi_{C}^{ij}+\sum_{L=1}^{\infty}(\frac{1}{\sqrt{h}}\Pi_{\tilde{B_{L}}}^{ij} -a^i \tilde{A}^{j}_{L} -a^j \tilde{A}^{i}_{L})-\frac{1}{\sqrt{h}}\Pi^{ij}_{\tilde{A}_1} +a^i \tilde{B}^{j}_{1} \\ \nonumber +a^j \tilde{A}^{i}_{1}\Big\}+N^{k}
\Big\{-\mathcal{D}_{k}\tilde{C}^{ij}+\mathcal{D}_{k}\tilde{B}^{ij}-\mathcal{D}_{k}\tilde{A}^{ij}\Big\}+ f_{2_{-0}}\lambda^{ij}_{\tilde{Q}}\Bigg]\approx 0.
\end{eqnarray}
By this constraint we determine the Lagrange multiplier. Similarly for observable $\mathcal{Q}_{i},$ we have,
\begin{eqnarray}
G^{i}_{\mathcal{Q}}=\dot{\Pi^{i}}_{\mathcal{Q}}=\{\Pi^{i}_{\mathcal{Q}},\mathcal{H}_{tot}\}=N\sqrt{h} \Big\{-4 f_{2_{-0}} \mathcal{Q}^{i}-\sum_{n=1}^{\infty}(f_{2_{-n}}\mathcal{A}^{i}_{n})+2\mathcal{B}^{i}\Big\}\approx 0,
\end{eqnarray}
and
\begin{equation}
\{\Pi^{i}_{\tilde{Q}},G^{i}_{\mathcal{Q}}\}\neq0.
\end{equation}
These relations indicate that $\Pi^{i}_{\tilde{Q}}$ and $G^{i}_{\mathcal{Q}}$ are second class constraints. The evolution of $G^{i}_{\mathcal{Q}}$ becomes,
\begin{eqnarray}
\dot{G^{i}_{\mathcal{Q}}}=\{G^{i}_{\mathcal{Q}},\mathcal{H}_{tot}\}=N\sqrt{h}\Bigg[N\Big\{\sqrt{h}f_{2_{-n}}(\frac{1}{\sqrt{h}}\Pi^{i}_{\mathcal{B}^{L}}-a_{i})-\frac{1}{2}(\frac{1}{\sqrt{h}}\Pi^{i}_{\mathcal{A}^{1}}-a_{i})\Big\} \nonumber\\+N^{k}\Big\{\mathcal{D}_{k}\tilde{B}_{L}^{i}-\frac{1}{2}\mathcal{D}_{k}\tilde{A}_{1}^{i}\Big\}+4f_{2_{-0}}\lambda_{\mathcal{Q}^{i}}\Bigg]\approx 0.
\end{eqnarray}
Similar expressions can be developed for observables $\theta$ and corresponding second class constraints. The second class constraints $\Pi_{\theta}$ and $G_{\theta}$ satisfy,
\begin{equation}
G_{\theta}=\{\Pi_{\theta},\mathcal{H}_{tot}\}=N\sqrt{h}\Big\{-\Omega-\Psi_{1}+2f_{2_{-0}} \theta-\sum_{n=1}^{\infty}f_{2_{-n}}\Lambda_{n}\Big\}\approx0,
\end{equation}
and
\begin{equation}
\{\Pi_{\theta},G_{\theta}\}\neq0,
\end{equation}
indicate that $ \Pi_{\theta}$ and $G_{\theta} $ are second class constraints.
The evolution of $G_{\theta}$ is determined to be,
\begin{eqnarray}
\dot{G_{\theta}}=\{G_{\theta},\mathcal{H}_{tot}\}=N\sqrt{h}\Bigg[N\Big\{\frac{2}{\sqrt{h}}\Omega\Pi_{\Omega}-\frac{h_{ij}\Pi^{ij}}{\Omega}-\frac{\Pi_{\lambda_{1}}}{\sqrt{h}}+\sum_{n=1}^{\infty}f_{2_{-n}}\frac{\Pi_{\Psi_{n}}}{\sqrt{h}}
\Big\}+\\N^{k}\Big\{\mathcal{D}_{k}\Psi_{1}-\mathcal{D}_{k}\Omega+\mathcal{D}_{k}\Lambda_{L}\Big\}+2f_{2_{-0}} \lambda_{\theta} \Bigg] \approx0.
\end{eqnarray}
The last constraint is determined for observable $C_{i}.$ It is given by
\begin{equation}
G^{i}_{C}=\dot{\Pi^{i}}_{C}=\{\Pi^{i}_{C},\mathcal{H}_{tot}\}\approx 0,
\end{equation}
and
\begin{equation}
\{ \Pi^{i}_{C},G^{i}_{C}\}=0,
\end{equation}
indicate that $ \Pi_{i}^{C}$ and $G^{i}_{C} $ are first class constraints.
and it's evolution becomes
\begin{equation}
\dot{G^{i}_{C}}=\{G^{i}_{C},\mathcal{H}_{tot}\}\approx 0.
\end{equation}
Now we analyse the structure and nature of constraints for  different cases. 
\section{Calculation for Different Value of $n$}
\subsection{ Calculation for n=0 in $ S_{01}$} 
As a first case by taking n=0 in eq. (\ref{s1}), we notice that all the $ f_{1_{-n}}$, $A_{L}$ and $B_{L}$  not appear in the action eq. (\ref{s1}) except $f_{1_{-0}}$,
\begin{equation}
\begin{split}
S_{01}^{eqv} & =  \int d^{3}x N\sqrt{h} \Bigg[ Q + f_{1_{-0}} Q^2  + C\Big( K_{ij}K^{ij} -K^{2} + \mathcal{R} -Q\Big) -2K \nabla_{n}C \Bigg].
\end{split}
\end{equation}
In this case the canonical momenta satisfy,
\begin{eqnarray}
 \Pi_{N}& \approx& 0, \qquad
 \Pi_{i}  \approx 0, \qquad 
 \Pi^{ij} = \sqrt{h} B( \mathcal{K}^{ij} -h^{ij}\mathcal{K} ) - \sqrt{h} h^{ij}\nabla_{n}B, \qquad \qquad \qquad \qquad  \nonumber \\
 P_{Q}&  \approx &  0 , \qquad \qquad
 P_{C} = -2\sqrt{h} \mathcal{K}.
\end{eqnarray}
Hence the primary constraints are summarised as,
\begin{eqnarray*}
\Pi_{N}\approx 0, \qquad\qquad \Pi_{i}\approx 0,\qquad \qquad P_{Q}\approx 0.
\end{eqnarray*}
Now the Hamiltonian density for $n=0$ case,
\begin{equation}
\mathcal{H}= N \mathcal{H}_{N} + N^{i}\mathcal{H}_{i},
\end{equation}
with
\begin{equation}
\begin{split}
\mathcal{H}_{N} & = \frac{1}{\sqrt{h} C} \Pi^{ij} h_{ik} h_{jl} \Pi^{kl} -\frac{1}{3\sqrt{h} C} \Pi^{2} -\frac{1}{3\sqrt{h}} \Pi P_{C} + \frac{1}{6 \sqrt{h}} C P_{C}^{2} -\sqrt{h} C \mathcal{R} \\& + \sqrt{h}C Q  - \sqrt{h} Q (1 + f_{1_{-0}}Q),   \label{hnso1}
\end{split}
\end{equation}
and
\begin{equation}
\begin{split}
\mathcal{H}_{i} = -2 h_{ik}\mathcal{D}_{j}\Pi^{kl} + P_{C}\partial_{i}C.    \label{hiso1}
\end{split}
\end{equation} 
The total Hamiltonian with primary constraints becomes,
\begin{equation}
H_{total} = \int d^{3}x \Big( N \mathcal{H}_{N} + N^{i}\mathcal{H}_{i}+\lambda^{i}\Pi_{i}+\lambda^{N}\Pi_{N} + \lambda^{Q}P_{Q} \Big).
\end{equation}
Secondary constraints
$H_N \approx 0$, $ H_i \approx 0$.
We are interested to find Poisson bracket between primary constraint $ P_{Q}$ and total Hamiltonian so,
\begin{eqnarray}
G_{Q} = \partial_{t}P_{Q} =\{ P_{Q}, \mathcal{H}_{tot}\} = N\sqrt{h}\Big\{-C +1 + 2f_{1_{-0}}Q\Big\}.
\end{eqnarray}
Then time evolution of $G_{Q}$ become
\begin{eqnarray}
\dot{G}_{Q} = \{ G_{Q},\mathcal{H}_{tot}\}= N \Bigg( \frac{N}{3}(\Pi -CP_{C})+ \sqrt{h}f_{0}\lambda^{Q} \Bigg),
\end{eqnarray}
 $P_Q$ and $ G_{Q} $ are second class constraint since 
\begin{equation}
\{P_Q,G_Q\}=\sqrt{h}Q\neq 0.
\end{equation}
We use the general formula to count the number of the physical degrees of freedom,
$$N=\frac{1}{2}(2A-B-2C),$$
In this case,
\begin{eqnarray}
 A &= &\Bigg\{(h_{ij},\Pi_{ij}),(N,\Pi_{N}),N^{i},\Pi_{i}, (C,P_{C}),(Q,P_{Q})\Bigg\},\nonumber\\
 C &=& \Bigg\{\Pi_N, \mathcal{H}_{N},\Pi_{i},\mathcal{H}_{i}\Bigg\},\nonumber\\
 B &=&\Bigg\{P_{Q},G_{Q}\Bigg\},
\end{eqnarray}
Where,
\begin{eqnarray}
2A &=& 2 \times ( 6 + 1 + 3 + 1+ 1)= 24  \nonumber\\
2C &=& 2 \times (  1 + 1 + 3 +3)= 16 \nonumber\\
B &=& 2 \nonumber
\end{eqnarray}
 so we obtain,
 \begin{equation}
 N=\frac{1}{2}(24-2-16)=3
\end{equation}
The calculation we perform in this section is similar for action of the form,
\begin{equation}
 S=\int d^{3}x N\sqrt{h} (R+R^2)
\end{equation}
The degree of freedom for this is 3, so we verified our result.
\subsection{Calculation for n=1 in $ S_{01}$} 
We repeat here the same analysis as we perform in previous subsection  for $n=1$. In  Eq.(\ref{s1}) in this case all $ A_{n}$, $B_{n} $ and $f_{1_{-n}}$   become zero except $ A_{1}$, $B_{1} $  and  $f_{1_{-1}}$ we write it as  $ A_1$,   $B_1 $ and  $f_{1}$, so the action reduced to,
\begin{eqnarray}
S_{01}^{eqv}= \int d^{3}x N\sqrt{h} \Bigg[ Q  + Q f_{1}A_{1} + C\Big( K_{ij}K^{ij} -K^{2} + \mathcal{R} -Q\Big)\nonumber\\ -2K \nabla_{n}C - B_{1}Q + h^{ij} \mathcal{D}_{i}B_{1}\mathcal{D}_{j}A_{1}- \nabla_{n}B_{1}\nabla_{n}A_{1}\Bigg].\qquad
\end{eqnarray}
We read off canonical momenta as,
\begin{eqnarray*}
\Pi_{N} \approx 0, \qquad\qquad
\Pi_{i} \approx 0, \qquad 
\Pi^{ij} = \sqrt{h} C( \mathcal{K}^{ij} -h^{ij}\mathcal{K} ) - \sqrt{h} h^{ij}\nabla_{n}C, \qquad \qquad \qquad \qquad   \\
P_{Q} \approx  0 , \qquad \qquad
P_{C} = -2\sqrt{h} \mathcal{K}, \qquad
P_{A_{1}} = \sqrt{h}  \nabla_{n} B_{1}, \qquad  \qquad
P_{B_{1}} = \sqrt{h} \nabla_{n} A_{1}. \qquad \qquad
\end{eqnarray*}
Hence the primary constraints are summarised as
\begin{eqnarray*}
\Pi_{N}\approx 0 , \Pi_{i}\approx 0 , P_{Q}\approx 0.
\end{eqnarray*}
Now the Hamiltonian density become,
\begin{equation}
\mathcal{H}= N \mathcal{H}_{N} + N^{i} \mathcal{H}_{i},
\end{equation}
with
\begin{equation}
\begin{split}
\mathcal{H}_{N} & = \frac{1}{\sqrt{h} C} \Pi^{ij} h_{ik} h_{jl} \Pi^{kl} -\frac{1}{3\sqrt{h} C} \Pi^{2} -\frac{1}{3\sqrt{h}} \Pi P_{C} + \frac{1}{6 \sqrt{h}} C P_{C}^{2} -\sqrt{h} C \mathcal{R} \\&
 + \sqrt{h}C Q -\Big(\sqrt{h} Q + Q f_{1}A_{1}\Big) +\sqrt{h} 
 B_{1}Q  \\&- \frac{P_{A_{1}}P_{B_{1}}}{\sqrt{h}} -\sqrt{h} h^{ij}  \mathcal{D}_{i}B_{1}\mathcal{D}_{j}A_{1}\Big),  \label{hnso1}
\end{split}
\end{equation}
and
\begin{equation}
\begin{split}
\mathcal{H}_{N} = -2 h_{ik}\mathcal{D}_{j}\Pi^{kl} +  P_{A_{1}}\mathcal{D}_{i}A_{1} + P_{B_{1}}\mathcal{D}_{i}B_{1}  + P_{C}\mathcal{D}_{i}C.    \label{hiso1}
\end{split}
\end{equation} 
The total Hamiltonian with primary constraints becomes,
\begin{equation}
H_{tot} = \int d^{3}x \Big( N \mathcal{H}_{N} + N^{i} \mathcal{H}_{i}+\lambda^{i}\Pi_{i}+\lambda^{N}\Pi_{N} + \lambda^{Q}P_{Q} \Big),
\end{equation}
the secondary constraints are:
$\mathcal{H}_{N} \approx 0$, $\mathcal{H}_{i} \approx 0$. 
Now again we check the Poisson bracket of $ P_Q $ with total Hamiltonian,
\begin{eqnarray}
G_{Q} = \partial_{t}P_{Q} =\{ P_{Q}, \mathcal{H}_{tot}\} = N\sqrt{h}\Big\{-C +(1 + f_{1} A_{1}) - B_{1}\Big\}.
\end{eqnarray}
Then time evolution of $G_{Q}$ become
\begin{eqnarray}
\dot{G}_{Q} = \{ G_{Q},\mathcal{H}_{tot}\}=
 N \Bigg[ \frac{N}{3}(\Pi -CP_{C}) + N f_{1}P_{B_{1}} 
+ P_{A_{1}}  \nonumber \\ + \sqrt{h}N^{i} \Big\{ \sum_{L=1}^{\infty}(\partial_{i}A_{1})f_{1} +(\partial_{i}B_{L})f_{L} \Big\}  + \sqrt{h}f_{0}\lambda^{Q} \Bigg]\approx0.
\end{eqnarray}
Whereas $P_Q$ is a first class constraint since 
\begin{equation}
\{P_Q,G_Q\}= 0.
\end{equation}
From the equation of motion for $ B_L $, we derive
\begin{equation}
A_1=\Box Q, 
\end{equation}
we obtain other primary constraints which are
\begin{eqnarray}
Z_{1} &=& A_{1} - \Box Q = 0. 
\end{eqnarray}
 Moreover $  A_{1} - \Box Q \approx 0 $   are primary constraints since the Poisson brackets of these quantities with total Hamiltonian  vanish weakly by use their equation of motion.\\
In this case,
\begin{eqnarray}
 A &= &\Bigg\{(h_{ij},\Pi_{ij}),(N,\Pi_{N}),N^{i},\Pi_{i}, (C,P_{C}),(Q,P_{Q}),(B_1,P_{B_1}),(A_{1},P_{A_{1}})\Bigg\},\nonumber\\
  C &=& \Bigg\{\Pi_N, \mathcal{H}_{N},\Pi_{i},\mathcal{H}_{i},P_{Q},G_{Q},Z_1\Bigg\}
\end{eqnarray}
Where,
\begin{eqnarray}
2A &=& 2 \times ( 6 + 1 + 3 + 1+ 1 +1+ 1)= 28 \nonumber\\
2C &=& 2 \times (  1 + 1 + 3 +3 1+ 1+1)= 22 \nonumber\\
B &=& 0 \nonumber\\
\end{eqnarray}
 so we obtain,
 \begin{equation}
 N=3
 \end{equation}
\subsection{Calculation for n=1 in $ S_{2}$}
In this section we develop the previous calculation for $ S_{2}^{eqv} $ for n=1\cite{Ferreira:2013tqn}. In this case all $ A^{ij}_{n}$, $B^{ij}_{n} $ and $f_{2_{-n}}$   become zero except $ A^{ij}_{1}$, $B^{ij}_{1} $  and  $f_{2_{-1}}$ we denote it as  $ A^{ij}$,   $B^{ij} $ and  $f_{2}$ then,
\begin{eqnarray}
S_{2}^{eqv} =\int d^{4}x \sqrt{-g}\Big[Q_{\mu \nu} f_{2}A^{\mu \nu}+I_{\mu \nu}(R^{\mu \nu} -Q^{\mu \nu})- B_{ \mu \nu}Q^{\mu \nu}+ B_{ \mu \nu}\Box A^{\mu \nu}\Big].
\end{eqnarray}
Now the action in 3+1 formalism is,
\begin{eqnarray}
S_{2}^{eqv}=\int d^{4}x N\sqrt{h}\Bigg[f_{2}\Big(\tilde{Q}_{ij}\tilde{A}^{ij}- 2\mathcal{Q}_{i}\mathcal{A}^{i}+\theta \Lambda\Big)+C^{ij}\Big(\mathcal{R}_{ij}-\tilde{Q}_{ij}\Big) +{\Omega}
\Big(- K_{s t} K^{s t}+K^{2}-\theta\Big)-\nonumber\\
K_{ij}\nabla_{\bold{n}}C^{ij}+K\nabla_{\bold{n}}\Omega +\tilde{B}_{ij}\tilde{Q}^{ij}-2\mathcal{B}_{i}\mathcal{Q}^{i}+\Psi \theta +\Big\{(D _{k}\tilde{B}_{ij}-K_{ki}\mathcal{B}_{ j}-\mathcal{B}_{ i} K_{kj})
(D ^{k}\tilde{A}^{i j}-\nonumber\\K^{ki}\mathcal{A}^{j}-\mathcal{A}^{i} K^{kj}) +(\nabla_{n}\tilde{B}_{ i j}+a_{i}\mathcal{B}_{ j}+a_{j}\mathcal{B}_{ i})(\nabla_{n}\tilde{A}^{i j}+a^{i}\mathcal{A}^{j}+a^{j}\mathcal{A}^{i}) +
4\Big( D_{k}\mathcal{B}_{j} - \Psi K_{kj}\Big)\nonumber\\ \Big(D ^{k}\mathcal{A}^{j}-\Lambda K^{kj}\Big)-4\Big(a_{j}-\nabla_{n}\mathcal{B}_{ j}\Big)\Big(a^{j}- \nabla^{n}\mathcal{A}^{j} \Big)+ 
D_{k}\Psi D^{k}\Lambda -\nabla_{n}\Psi \nabla_{n}\Lambda \Big\}\Bigg].
\end{eqnarray}
The primary constraints are,
$$\Pi^{ij}_{\tilde{Q}}\approx0, \Pi^{i}_{\tilde{Q}}\approx0,
\Pi_{\theta}\approx0,\Pi^{i}_{C}\approx0, N\approx0, N^{i}\approx0.$$
Now the Hamiltonian density becomes,
\begin{equation}
\mathcal{H}= N \mathcal{H}_{N} + N^{i} \mathcal{H}_{i},
\end{equation}
with
\begin{equation}
\begin{aligned}
\mathcal{H}_{N} & = \sqrt{h} \Big\{-f_{2}\Big(\tilde{Q}_{ij}\tilde{A}^{ij}- 2\mathcal{Q}_{i}\mathcal{A}^{i}+\theta \Lambda\Big) -\tilde{C}^{ij}\Big(\mathcal{R}_{ij} -\mathcal{Q}_{ij}\Big) -\Omega ( \Pi_{\Omega}^{2}- h_{sp}h_{tp}\Pi^{pq}_{\tilde{C}} \Pi^{st}_{\tilde{C}}-\theta)  +\\& \tilde{B}_{ij}\tilde{Q}^{ij} -2\mathcal{B}_{i}\mathcal{Q}^{i}+
\Psi \theta \Big\}
 + \frac{\Pi^{ij}_{h}}{\sqrt{h} \Omega} ( h_{ki}h_{lj}\Pi_{\tilde{C}}^{kl} + h_{ij} \Pi_{\Omega}) 
+ \Pi^{ij}_{\tilde{A}}\Big\{ \frac{1}{\sqrt{h}} h_{ki}h_{lj}\Pi_{\tilde{B}}^{kl} \\&-a_i \mathcal{A}_{j} -a_j \mathcal{A}_{i} \Big\} + \Pi^{ij}_{\tilde{B}}\Big\{ \frac{1}{\sqrt{h}} h_{ki}h_{lj}\Pi_{\tilde{A}}^{kl} -a_i \mathcal{B}_{j}  -a_j \mathcal{B}_{i} \Big\} 
 -\frac{\Pi^{i}_{\mathcal{A}}}{4} \Big\{ \frac{1}{\sqrt{h}} h_{ki}\Pi_{\mathcal{B}}^{k} -a_i \Big\}
 \\&
  -\frac{\Pi^{i}_{\mathcal{B}}}{4} \Big\{ \frac{1}{\sqrt{h}} h_{ki}\Pi_{\mathcal{A}}^{k} -a_i \Big\} + 2 \frac{1}{\sqrt{h}} \Pi_{\Psi}\Pi_{\Lambda} + 
\Big[ \Big\{ \mathcal{D}_{s}\tilde{B}_{ij} + \frac{1}{\sqrt{h}} h_{s k}h_{il} \Pi^{kl}_{\tilde{C}}  \tilde{B}_{j} 
 \\&
+ \frac{1}{\sqrt{h}} h_{s k}h_{jl} \Pi^{kl}_{\tilde{C}}  \tilde{B}_{i} \Big\} 
 \Big\{ \mathcal{D}^{s}\tilde{A}^{ij} + \frac{1}{\sqrt{h}}\Pi^{s i}_{\tilde{C}}\mathcal{A}^{j} + \frac{1}{\sqrt{h}} \mathcal{A}^{i}\Pi^{sj}_{\tilde{C}}\Big\}  + \frac{1}{\sqrt{h}} h_{ki}h_{lj}\Pi_{\tilde{B}}^{kl}\Pi^{ij}_{\tilde{A}} \\&
  + 4 \sqrt{h}\Big\{\mathcal{D}_{s} \mathcal{B}_{j} + \frac{1}{\sqrt{h}}\Psi h_{s k}h_{jl} \Pi^{kl}_{\tilde{C}}\Big\}  \Big\{\mathcal{D}^{s}\mathcal{A}^{j} + \frac{1}{\sqrt{h}} \Lambda \Pi^{s j}_{\tilde{C}} \Big\} -\frac{1}{4 \sqrt{h}} \Pi^{i}_{\tilde{A}} h_{ik} \Pi^{k}_{\tilde{B}} -\\& \sqrt{h} \mathcal{D}_{s} \Psi \mathcal{D}^{s} \Lambda + \frac{1}{\sqrt{h}} \Pi_{\Psi}  \Pi_{\Lambda}    \Big]
\end{aligned},
\end{equation}
and
\begin{equation}
\begin{aligned}
\mathcal{H}_{k}  &= \Bigg[ \Pi^{ij}_{\tilde{A}} \mathcal{D}_{k}\tilde{B}_{ij}  + \Pi^{ij}_{\tilde{B}} \mathcal{D}_{k}\tilde{A}_{ij}  -\frac{1}{4}\mathcal{D}_{k}\tilde{B}_i \Pi^{i}_{\tilde{A}}  -\frac{1}{4}\mathcal{D}_{k}\tilde{A}_i \Pi^{i}_{\tilde{B}} + \Pi_{\Psi} \mathcal{D}_{k} \Psi+ \Pi_{\Lambda} \mathcal{D}_{k} \Lambda  \\& + \Pi_{\tilde{C}_{ij}} \mathcal{D}_{k} \tilde{C}^{ij} - \Pi_{\Omega} \mathcal{D}_{k}\Omega  \Bigg].
\end{aligned}
\end{equation}
From the Poisson brackets,
\begin{equation}
G^{ij}_{\tilde{Q}}=\dot{\Pi^{ij}}_{\tilde{Q}}=\{\Pi^{ij}_{\tilde{Q}},\mathcal{H}_{tot}\}=N\sqrt{h}\Big\{ f_{2}\tilde{A}^{ij}-\tilde{\tilde{C}}^{ij}- \tilde{B}^{ij}\Big\}\approx 0,
\end{equation}
and
\begin{equation}
\{\Pi^{ij}_{\tilde{Q}},G^{ij}_{\tilde{Q}}\}=0 ,
\end{equation} suggest that $\Pi^{ij}_{\tilde{Q}}, G^{ij}_{\tilde{Q}}$ are first class constraint.\\
From the condition,
\begin{equation}
\dot{G^{ij}}_{\tilde{Q}}=\{G^{ij}_{\tilde{Q}},\mathcal{H}_{tot}\}\approx 0,
\end{equation}
\begin{eqnarray}
=N\sqrt{h}\Bigg[N\Big\{-2\sqrt{h}\Pi^{ij}_{\tilde{C}}-\frac{\Pi^{ij}_{h}}{\sqrt{h} \Omega} +\frac{4}{\sqrt{h}} \Pi^{ki}_{\tilde{C}}(\mathcal{A}^{j}\mathcal{B}_{k}+\mathcal{A}_{k}\mathcal{B}^{j})+2\mathcal{B}^{k}\mathcal{D}^{j}\mathcal{A}^{i}_{k}+2\mathcal{A}^{k}\mathcal{D}^{j}\mathcal{B}^{i}_{k} \nonumber\\ +4\mathcal{D}^{i}\mathcal{A}^{j}(\Psi +\Lambda )+\frac{2}{\sqrt{h}}\Psi \Lambda \Pi_{\tilde{C}}^{ij}+\frac{1}{\sqrt{h}}\Pi_{\tilde{B}}^{ij} -a^i \tilde{A}^{j} -a^j \tilde{A}^{i})-\frac{1}{\sqrt{h}}\Pi^{ij}_{\tilde{A}} +a^i \tilde{B}^{j} \\ \nonumber +a^j \tilde{A}^{i}\Big\}+N^{k}
\Big\{-\mathcal{D}_{k}\tilde{\tilde{C}}^{ij}+\mathcal{D}_{k}\tilde{B}^{ij}-\mathcal{D}_{k}\tilde{A}^{ij}\Big\}+ f_{2_{-0}}\lambda^{ij}_{\tilde{Q}}\Bigg]\approx 0.
\end{eqnarray}
 we can find the value of Lagrange multiplier $ \lambda_{\tilde{Q}^{ij}} $.
 Again from the condition,
 \begin{equation}
G^{i}_{\mathcal{Q}}=\dot{\Pi^{i}}_{\mathcal{Q}}=\{\Pi^{i}_{\mathcal{Q}},\mathcal{H}_{tot}\}=N\sqrt{h} \Big\{-f_{2}\mathcal{A}^{i}+2\mathcal{B}^{i}\Big\}\approx 0,
\end{equation}
and
\begin{equation}
\{\Pi^{i}_{\tilde{Q}},G^{i}_{\mathcal{Q}}\}=0,
\end{equation}
we obtain  $ \Pi^{i}_{\tilde{Q}}$ and $G^{i}_{\mathcal{Q}} $  as first class constraint.
Similarly from the relation,
\begin{eqnarray}
\dot{G^{i}_{\mathcal{Q}}}=\{G^{i}_{\mathcal{Q}},\mathcal{H}_{tot}\}=N\sqrt{h}\Bigg[N\Big\{\sqrt{h}f_{2}(\frac{1}{\sqrt{h}}\Pi^{i}_{\mathcal{B}}-a_{i})-\frac{1}{2}(\frac{1}{\sqrt{h}}\Pi^{i}_{\mathcal{A}}-a_{i})\Big\} \nonumber\\+N^{k}\Big\{\mathcal{D}_{k}\tilde{B}^{i}-\frac{1}{2}\mathcal{D}_{k}\tilde{A}^{i}\Big\}+4f_{2_{-0}}\lambda_{\mathcal{Q}^{i}}\Bigg]\approx 0.
\end{eqnarray}
 we can find the value of Lagrange multiplier $ \lambda_{\mathcal{Q}^{i}} $.
 For canonical variable $ \theta $, we have
\begin{equation}
G_{\theta}=\{\Pi_{\theta},\mathcal{H}_{tot}\}=N\sqrt{h}\Big\{-\Omega-\Psi-f_{2}\Lambda\Big\}\approx0,
\end{equation}
and
\begin{equation}
\{\Pi_{\theta},G_{\theta}\}=0,
\end{equation}
 $\Pi_{\theta}$ and $G_{\theta}$ are first class constraint.
 By the relation,
\begin{equation}
\dot{G_{\theta}}=\{G_{\theta},\mathcal{H}_{tot}\}=N\sqrt{h}\Bigg[N\Big\{\frac{2}{\sqrt{h}}\Omega\Pi_{\Omega}-\frac{h_{ij}\Pi^{ij}}{\Omega}-\frac{\Pi_{\lambda}}{\sqrt{h}}+f_{2}\frac{\Pi_{\Psi}}{\sqrt{h}}
\Big\}+\\N^{k}\Big\{\mathcal{D}_{k}\Psi-\mathcal{D}_{k}\Omega+\mathcal{D}_{k}\Lambda\Big\}+2f_{2} \lambda_{\theta} \Bigg] \approx0
\end{equation}
we can find the value of Lagrange multiplier $ \lambda_{\theta} $.\\
For canonical varible $ C^{i} $
\begin{equation}
G^{i}_{C}=\dot{\Pi^{i}}_{C}=\{\Pi^{i}_{C},\mathcal{H}_{tot}\}\approx 0
\end{equation}
\begin{equation}
\dot{G^{i}_{C}}=\{G^{i}_{C},\mathcal{H}_{tot}\}\approx 0
\end{equation}
From the 
equations of motion of $B_{\mu \nu}  $, we obtain another  primary constraint as,
\begin{eqnarray}
Z^{\mu\nu}=Q^{\mu \nu}- \Box A^{\mu \nu}\approx0.
\end{eqnarray}
After identifying all the primary and secondary constraint and further dividing it first and second class constraint. We now count the number of physical degree of freedom for $ S_2 $.
 We have 44 first class constraint out of   50 total variable. Now from
$$N=\frac{1}{2}(2A-B-2C),$$
In this case,
$$A=50 ,\qquad B=0 \qquad C=44 $$
like the previous analysis for action $ S_{01} $ for $ n=0 $ and $ n=1$, in this case we get six physical degree of freedom.
\section{Conclusion}
In this paper, we have presented Hamiltonian formalism of a class of nonlocal theory generally known as quadratic non local theory, in which action ( eq.1) contain inverse d'Alembertian  operators acting on  curvature term. and it is characterized by a function $ \mathcal{F}(\Box) $, in our case its form is $ \frac{1}{\Box^{n}} $,  In order to simply our calculation, first we split action into four parts as $ S_{0}$ ( eq.2), $S_{1} $ ( eq.3),  $S_{2} $ ( eq.4)  and $ S_{4} $ ( eq.5). Our present work is focussed on action $ S_{0}$, $S_{1} $(together we call $ S_{01} $) and  $S_{2} $.

   After expressing nonlocal action in the form of equivalent scalar  tensor action we construct Hamiltonian for both cases ($ S_{01} $ and  $S_{2} $). The Hamiltonian is linear in shift ($ N $) and lapse ($ N^{i} $), enabling us to identify all constraints associated with the canonical variable. Here we have calculating all the Poisson bracket of canonical variables and using Poisson algebra we classify into the first and second class constraint. There is  infinite number of variable appear in for $ S_{01} $ and  $S_{2} $ so we get infinite number of degree of freedoms(dof's). The constraint we have found here is consistent with the theory.  
   
    Next we perform Hamiltonian analysis for  $ n=0 $ in $ S_{01} $, where action contain  $ R+R^{2} $ term as Lagrangian density. In this case, we obtain 8 first-class constraints and 2   second class constraints  out of 24 total phase space variable, which amounts to 3 dof's \\
    Again we perform the similar calculation for  n=1 in $ S_{01} $, we obtain $ R+R\frac{1}{\Box}R $ as Lagrangian density, here we get 11 first  class constraint and no second class constraint out of 28 total phase space variable, which in turns gives rise to 3 dof's.
     
     As a final case  we calculate the dof's for $ R_{\mu\nu}\frac{1}{\Box}R^{\mu\nu} $ by putting $ n=1  $ in $ S_2 $. Here we get 22 first  class constraint and no second class constraint out of 50 phase space variable. This case having more complex tensor calculation due to the tensor nature of constraints. We calculate all the Poisson bracket necessary for classifying the constraints and finding the  dof's. In this case, we have 6 dof's
    
This analysis can be generalized to find the number of degree of freedom for a more complicated form of our nonlocal action with Ricci scalar and Ricci tensor. The detailed analysis of the nature of the physical degree of freedom will be reported elsewhere.

\section{ACKNOWLEDGEMENT}
This work was partially funded by DST (Govt. of India), Grant No. SERB/PHY/2017041. Calculations were
performed using xAct packages of Mathematica.
\section*{Appendix: A}
In this appendix, we write a simplified  expression for higher derivative terms obtain in eq.(37), the first higher derivative term is,
\begin{eqnarray}
L_{H1}= \int d^{4}x N\sqrt{h}C^{ij}L_{n}  K_{ij}.
\end{eqnarray}
Now, the covariant form of  $L_{n}  K_{ij}$,
\begin{equation}
L_{n}  K_{ij}=n^{a}\nabla_{a}K_{ij}+K_{aj}\nabla_{i}n^{a}+K_{ai}\nabla_{j}n^{a}\label{ki}
\end{equation}
By using eq.\ref{ki}
\begin{eqnarray}
L_{H1} & = &\int d^{4}x N\sqrt{h}C^{ij}\Big(n^{a}\nabla_{a}K_{ij}+K_{aj}\nabla_{i}n^{a}+K_{ai}\nabla_{j}n^{a}\Big)
\end{eqnarray}
where, $$K_{ij}=\nabla_{i}n_{j}+n_{i}a_{j}$$ $$ K=\nabla_{i}n^{i}$$
After further simplification, we get,
where, we have used the relation  $n_{j} C^{ij}=0$ which gives us,
\begin{eqnarray}
L_{H1}=\int d^{4}x N\sqrt{h}[-K_{ij}KC^{ij}-n^{a}(\nabla_{a}C^{ij})K_{ij}+2K_{ai}K^{a}_{j}C^{ij}]
\end{eqnarray}
We perform a similar calculation for $L_{H2}$
\begin{eqnarray}
L_{H2}= \int d^{4}x N\sqrt{h}[- h^{s t} L_{n}  K_{s t}]\Omega.
\end{eqnarray}
The covariant form of  $h^{s t} L_{n}  K_{s t}$,
\begin{equation}
h^{s t} L_{n}  K_{s t}=n^{a}\nabla_{a}K+2K_{t}^{s}\nabla_{s}n^{t}\label{kii}
\end{equation}
By using eq.\ref{kii} and further simplification, it reduces to,
\begin{eqnarray}
L_{H2} & = & \int d^{4}x N\sqrt{h}[-\Omega K^{2}-(\nabla_{a}\Omega)n^{a}K+2K_{t}^{s} K_{s}^{t}\Omega]
\end{eqnarray}
\section*{Appendix: B}
We derive the expression for first covariant derivative of $ B_{\mu\nu} $ by using 3+1 decomposition,first we split$ B_{\mu\nu} $,
\begin{eqnarray}
B_{L \mu \nu} = \tilde{B}_{L \mu \nu}-n_{\mu}\mathcal{B}_{L \nu}-n_{\nu}\mathcal{B}_{L \mu}+n_{\mu}n_{\nu}\Psi
\end{eqnarray}
Now, covariant derivative of $ B_{\mu\nu} $ is,
\begin{eqnarray}
\nabla _{\rho}B_{L \mu \nu}  & = & \nabla _{\rho}(\tilde{B}_{L \mu \nu}-n_{\mu}\mathcal{B}_{L \nu}-n_{\nu}\mathcal{B}_{L \mu}+n_{\mu}n_{\nu}\Psi)
\end{eqnarray}
Expanding all the terms, it becomes
\begin{eqnarray}
\nabla _{\rho}B_{L \mu \nu} & = &\nabla _{\rho}\tilde{B}_{L \mu \nu}-\nabla _{\rho}n_{\mu}\mathcal{B}_{L \nu}-\nabla _{\rho}\mathcal{B}_{L \nu} n_{\mu}-\nabla _{\rho}n_{\nu}\mathcal{B}_{L \mu}-\nabla _{\rho}\mathcal{B}_{L \mu} n_{\nu} +n_{\mu}\Psi\nabla _{\rho}n_{\nu}\nonumber \\ & + &n_{\nu}\Psi\nabla _{\rho}n_{\mu}+n_{\mu}n_{\nu}\nabla_{\rho}\Psi \qquad
\end{eqnarray}
Further use $\nabla _{\rho}n_{\mu}=K_{\rho\mu}-n_{\rho}a_{\mu}$ and split $ \nabla_\mu=h^{\rho}_{\mu}\mathcal{D}_{\rho}-n_{\mu}\nabla_n $, and after rearranging, it takes the form,
\begin{eqnarray}
 \nabla _{\rho}B_{L \mu \nu} & = & D _{\rho}\tilde{B}_{L \mu \nu}-K_{\rho\mu}\mathcal{B}_{L \nu}-\mathcal{B}_{L \mu} K_{\rho\nu}-n _{\rho}\Big[\nabla_{n}\tilde{B}_{L \mu \nu}-a_{\mu}\mathcal{B}_{L \nu}-a_{\nu}\mathcal{B}_{L \mu}\Big]- n_{\mu} \Big[D _{\rho}\mathcal{B}_{L \nu}-\Psi K_{\rho\nu}\Big] \nonumber \\ & - &n_{\nu}\Big[D _{\rho}\mathcal{B}_{L \mu} -\Psi K_{\rho\mu}\Big]-n_{\mu}n_{\rho}\Big(a_{\nu}-\nabla_{n}\mathcal{B}_{L \nu}\Big)-n_{\nu}n_{\rho}\Big(a_{\mu}-\nabla_{n}\mathcal{B}_{L \nu}\Big)+n_{\mu}n_{\nu}\nabla_{\rho}\Psi
\end{eqnarray}
\section*{Appendix: C} Here we provide all the first derivative of all conjugate coordinates of $ S_{2} $.
\begin{eqnarray}
\dot{h}_{ij} & = & \frac{N}{\sqrt{h}\Omega}(h_{ki}h_{lj}\Pi_{C}^{kl} + h_{ij} \Pi_{\Omega} ) \\
\dot{\tilde{B}}_{L_{ij}} & = & \Bigg[ N \Big\{ \frac{1}{\sqrt{h}} (h_{ki}h_{lj}\Pi_{\tilde{A_{L}}}^{kl} -a_i \mathcal{B}_{L_{j}}  -a_j \mathcal{B}_{L_{i}} \Big\}  + N^{k} \mathcal{D}_k \tilde{B}_{L_{ij}}\Bigg] \\
\dot{\tilde{A}}_{L_{ij}} & = & \Bigg[ N \Big\{ \frac{1}{\sqrt{h}} (h_{ki}h_{lj}\Pi_{\tilde{B_{L}}}^{kl} -a_i \mathcal{A}_{L_{j}}  -a_j \mathcal{A}_{L_{i}} \Big\}  + N^{k} \mathcal{D}_k \tilde{A}_{L_{ij}}\Bigg] \\
\dot{\tilde{B}}_{L_{i}} & = & \Bigg[ \frac{N}{4} \Big\{ \frac{1}{\sqrt{h}} (h_{ki}\Pi_{\mathcal{A}_{L}}^{k} -a_i \Big\}  + N^{k} \mathcal{D}_k \mathcal{B}_{L_{i}}\Bigg] \\
\dot{\tilde{A}}_{L_{i}} & = & \Bigg[ \frac{N}{4} \Big\{ \frac{1}{\sqrt{h}} (h_{ki}\Pi_{\mathcal{B}_{L}}^{k} -a_i \Big\}  + N^{k} \mathcal{D}_k \mathcal{A}_{L_{i}}\Bigg] \\
\dot{\Psi} &=& \frac{N}{\sqrt{h}} \Pi_{\Lambda_{L}} + N^{k} \mathcal{D}_{k}\Psi_{L} \\
\dot{\Lambda} &=& \frac{N}{\sqrt{h}} \Pi_{\Psi_{L}} + N^{k} \mathcal{D}_{k}\Lambda_{L}
\end{eqnarray}


\begin{thebibliography}{0}%
\makeatletter
\providecommand \@ifxundefined [1]{%
 \@ifx{#1\undefined}
}%
\providecommand \@ifnum [1]{%
 \ifnum #1\expandafter \@firstoftwo
 \else \expandafter \@secondoftwo
 \fi
}%
\providecommand \@ifx [1]{%
 \ifx #1\expandafter \@firstoftwo
 \else \expandafter \@secondoftwo
 \fi
}%
\providecommand \natexlab [1]{#1}%
\providecommand \enquote  [1]{``#1''}%
\providecommand \bibnamefont  [1]{#1}%
\providecommand \bibfnamefont [1]{#1}%
\providecommand \citenamefont [1]{#1}%
\providecommand \href@noop [0]{\@secondoftwo}%
\providecommand \href [0]{\begingroup \@sanitize@url \@href}%
\providecommand \@href[1]{\@@startlink{#1}\@@href}%
\providecommand \@@href[1]{\endgroup#1\@@endlink}%
\providecommand \@sanitize@url [0]{\catcode `\\12\catcode `\$12\catcode
  `\&12\catcode `\#12\catcode `\^12\catcode `\_12\catcode `\%12\relax}%
\providecommand \@@startlink[1]{}%
\providecommand \@@endlink[0]{}%
\providecommand \url  [0]{\begingroup\@sanitize@url \@url }%
\providecommand \@url [1]{\endgroup\@href {#1}{\urlprefix }}%
\providecommand \urlprefix  [0]{URL }%
\providecommand \Eprint [0]{\href }%
\providecommand \doibase [0]{http://dx.doi.org/}%
\providecommand \selectlanguage [0]{\@gobble}%
\providecommand \bibinfo  [0]{\@secondoftwo}%
\providecommand \bibfield  [0]{\@secondoftwo}%
\providecommand \translation [1]{[#1]}%
\providecommand \BibitemOpen [0]{}%
\providecommand \bibitemStop [0]{}%
\providecommand \bibitemNoStop [0]{.\EOS\space}%
\providecommand \EOS [0]{\spacefactor3000\relax}%
\providecommand \BibitemShut  [1]{\csname bibitem#1\endcsname}%
\let\auto@bib@innerbib\@empty
\end{thebibliography}%


\begin{thebibliography}{99}
\bibitem{RevModPhys.61.1}
S. Weinberg, "The cosmological constant problem," Rev.\ Mod.\ Phys. {\bf 61}, 1, 1 (1989)

\bibitem{Wetterich:1997bz}
Wetterich. Christof, "Effective nonlocal Euclidean gravity, Gen.\ Rel.\ Grav .{\bf 30}, (1998).
      doi:10.1023/A:1018837319976,

\bibitem{Deser:2007jk}
Deser. Stanley and Woodard. R. P., "{Nonlocal Cosmology}", \ Phys. \ Rev. \ Lett., {\bf 99}, 2007,
doi :10.1103/PhysRevLett.99.111301, eprint:0706.2151,
      
\bibitem{Barvinsky:2002uf}
Barvinsky. A. O. and Mukhanov. Viatcheslav F., "New nonlocal effective action", \ Phys.\ Rev., {\bf D66},(2002),
doi:10.1103/PhysRevD.66.065007, eprint: hep-th/0203132,

\bibitem{Barvinsky:2011hd} 
Barvinsky A. O., Dark energy and dark matter from nonlocal ghost-free gravity theory, \ Phys. \ Lett., {\bf B710}, (2012),
    doi: 10.1016/j.physletb.2012.02.075, eprint :1107.1463,
      
\bibitem{Park:2017zls} 
Park, Sohyun, "Revival of the Deser-Woodard nonlocal gravity model:Comparison of the original nonlocal form and a localized formulation", \ Phys. \ Rev., {\bf D97},(2018),
   doi:10.1103/PhysRevD.97.044006,
 \bibitem{Kumar:2018pkb} 
  U.~Kumar, S.~Panda and A.~Patel,
  Phys.\ Rev.\ D {\bf 98}, no. 12, 124040 (2018)
  doi:10.1103/PhysRevD.98.124040
  [arXiv:1808.04569 [gr-qc]].
  
  \bibitem{Kumar:2019uwi} 
  U.~Kumar, S.~Panda and A.~Patel,
  arXiv:1906.11714 [gr-qc].
  
  \bibitem{Kumar:2019lzp} 
  U.~Kumar, S.~Panda and A.~Patel,
  arXiv:1908.08188 [gr-qc].   
\bibitem{Label1} 
M. Ostrogradski, Memoires sur les equations differentielles relatives au probleme des isoperimetres, Mem. Ac. St. Petersbourg VI (1850) 385.   
     
\bibitem{Stelle:1976gc} 
  K.~S.~Stelle,
  ``Renormalization of Higher Derivative Quantum Gravity,''
  Phys.\ Rev.\ D {\bf 16}, 953 (1977).
  doi:10.1103/PhysRevD.16.953


\bibitem{VanNieuwenhuizen:1973fi} 
  P.~Van Nieuwenhuizen,
  ``On ghost-free tensor lagrangians and linearized gravitation,''
  Nucl.\ Phys.\ B {\bf 60}, 478 (1973).
  doi:10.1016/0550-3213(73)90194-6


\bibitem{Biswas:2011ar} 
  T.~Biswas, E.~Gerwick, T.~Koivisto and A.~Mazumdar,
  ``Towards singularity and ghost free theories of gravity,''
  Phys.\ Rev.\ Lett.\  {\bf 108}, 031101 (2012)
  doi:10.1103/PhysRevLett.108.031101
  [arXiv:1110.5249 [gr-qc]].
  
\bibitem{Mazumdar:2017kxr} Talaganis, Spyridon and Teimouri, Ali, "{Hamiltonian Analysis for Infinite Derivative Field Theories and Gravity}", (2017) eprint: 1701.01009,
      
 \bibitem{Biswas:2013kla} 
  T.~Biswas, T.~Koivisto and A.~Mazumdar,
  ``Nonlocal theories of gravity: the flat space propagator,''
  arXiv:1302.0532 [gr-qc].

\bibitem{Wipf:1993xg} 
A.~W.~Wipf,
``Hamilton's formalism for systems with constraints,''
Lect.\ Notes Phys.\  {\bf 434}, 22 (1994)
doi:10.1007/3-540-58339-4\_14
[hep-th/9312078].
  
\bibitem{Dirac:1958sq} 
  P.~A.~M.~Dirac,
  Proc.\ Roy.\ Soc.\ Lond.\ A {\bf 246}, 326 (1958).
  doi:10.1098/rspa.1958.0141
  
  \bibitem{Dirac:1958sc} 
  P.~A.~M.~Dirac,
  Proc.\ Roy.\ Soc.\ Lond.\ A {\bf 246}, 333 (1958).
  doi:10.1098/rspa.1958.0142
  
\bibitem{Dirac1}
Dirac, P.A.M.,  ``Lectures on Quantum Mechanics'', Belfer Graduate School of Science, Yeshiva University, New York, (1964).
  
  \bibitem{Matschull:1996up} 
  H.~J.~Matschull,
  ``Dirac's canonical quantization program,''
  quant-ph/9606031.

\bibitem{Biswas:2013cha} 
T.~Biswas, A.~Conroy, A.~S.~Koshelev and A.~Mazumdar,
``Generalized ghost-free quadratic curvature gravity,''
 Class.\ Quant.\ Grav.\  {\bf 31}, 015022 (2014)
  Erratum: [Class.\ Quant.\ Grav.\  {\bf 31}, 159501 (2014)]
  doi:10.1088/0264-9381/31/1/015022, 10.1088/0264-9381/31/15/159501
  [arXiv:1308.2319 [hep-th]].
  
   \bibitem{Biswas:2016etb} 
  T.~Biswas, A.~S.~Koshelev and A.~Mazumdar,
  ``Gravitational theories with stable (anti-)de Sitter backgrounds,''
  Fundam.\ Theor.\ Phys.\  {\bf 183}, 97 (2016)
  doi:10.1007/978-3-319-31299-6
  [arXiv:1602.08475 [hep-th]].
  
  
  \bibitem{Biswas:2016egy} 
  T.~Biswas, A.~S.~Koshelev and A.~Mazumdar,
  ``Consistent Higher Derivative Gravitational theories with stable de Sitter and Anti-de Sitter Backgrounds,''
  arXiv:1606.01250 [gr-qc].
  

 
\bibitem{Biswas:2012bp}
  T.~Biswas, A.~S.~Koshelev, A.~Mazumdar and S.~Y.~Vernov,
  ``Stable bounce and inflation in nonlocal higher derivative cosmology,''
  JCAP {\bf 1208} (2012) 024
  doi:10.1088/1475-7516/2012/08/024
  [arXiv:1206.6374 [astro-ph.CO]].
 
\bibitem{Talaganis:2014ida} 
  S.~Talaganis, T.~Biswas and A.~Mazumdar,
  ``Towards understanding the ultraviolet behavior of quantum loops in infinite-derivative theories of gravity,''
  Class.\ Quant.\ Grav.\  {\bf 32}, no. 21, 215017 (2015)
  doi:10.1088/0264-9381/32/21/215017
  [arXiv:1412.3467 [hep-th]].
  
\bibitem{Kluson:2011tb} 
  J.~Kluson,
  ``Non-Local Gravity from Hamiltonian Point of View,''
  JHEP {\bf 1109}, 001 (2011)
  doi:10.1007/JHEP09(2011)001
  [arXiv:1105.6056 [hep-th]].
  
\bibitem{Arnowitt:1962hi} 
  R.~L.~Arnowitt, S.~Deser and C.~W.~Misner,
  ``The Dynamics of general relativity,''
  Gen.\ Rel.\ Grav.\  {\bf 40}, 1997 (2008)
  doi:10.1007/s10714-008-0661-1
  [gr-qc/0405109].
  
\bibitem{Gourgoulhon:2007ue} 
  E.~Gourgoulhon,
  ``$3+1$ formalism and bases of numerical relativity,''
  gr-qc/0703035 [gr-qc].
  

  
  
\bibitem{PhysRevD.88.123502}
A few cosmological implications of tensor nonlocalities,
 Ferreira, Pedro G. and Maroto, Antonio L., \ Phys. \ Rev. \ D, {\bf 88}, 2013, 10.1103/PhysRevD.88.123502,
 
 
\bibitem{PhysRevD.95.043539}
 Instabilities in tensorial nonlocal gravity,
Nersisyan, Henrik and Akrami, Yashar and Amendola, Luca and Koivisto, Tomi S. and Rubio, Javier and Solomon, Adam R., \ Phys. \ Rev. \ D, {\bf 95}, 2017, 10.1103/PhysRevD.95.043539,
 
 
\bibitem{PhysRevD.98.084040}
Scalar-tensor nonlocal gravity,
Tian, Shuxun, \ Phys. \ Rev. \ D, {\bf 98}, 2018, 10.1103/PhysRevD.98.084040,
 
 
\bibitem{Teimouri:2018ogt} 
  A.~Teimouri,
  doi:10.17635/lancaster/thesis/223
  arXiv:1811.09818 [gr-qc].
  
\bibitem{Ferreira:2013tqn}
    Ferreira, Pedro G. and Maroto, Antonio L.,
      ``A few cosmological implications of tensor nonlocalities,
   Gen. \ Phys. \ Rev.\ {\bf D88} 2013, 12, 10.1103/PhysRevD.88.123502,
\end{thebibliography}

\end{document}